\def\OPTIONArxiv{0}
\def\OPTIONLoudLabels{0}
\def\OPTIONArxiv{1}

\gdef\OPTIONArxiv{1}

\def\OPTIONConf{3}
\ifnum\OPTIONConf=3%
  \ifnum\OPTIONArxiv=0%
     \documentclass[format=acmsmall, review=true, screen=true, printccs=true]{acmart}
  \else
     \documentclass[format=acmsmall, review=false, screen=true]{acmart}
  \fi
\usepackage{graphics,pstricks,color}

\def~{\nobreakspace{}}

\usepackage{listings}

\usepackage{pgf,tikz}
\usetikzlibrary{shapes,arrows,matrix,fit,backgrounds,calc,decorations,decorations.pathmorphing}

\usepackage{xspace}

\usepackage{multirow}
\usepackage{stmaryrd}
\usepackage{pifont}
\usepackage{comment}
\usepackage{mathpartir}
\usepackage{framed}
\usepackage{rotating}

\usepackage{jdunfield}
\usepackage{llproof}
\usepackage{rulelinks}
\usepackage[OT2,T1]{fontenc}

\setcitestyle{nosort}    %

\ifnum\OPTIONConf=3%
\else%
   \usepackage[headings]{fullpage}
\fi

\usepackage{relsize}
\usepackage{booktabs}
\usepackage{breakurl}
\usepackage{latexsym}
\usepackage{stmaryrd}
\usepackage{amsmath,amsthm} %
\usepackage{thmtools,thm-restate}

\usepackage{enumerate}

\let\MathRightArrow\Rightarrow %
\usepackage{marvosym} %
\def\Rightarrow{\MathRightArrow}

\def\MathparLineskip{\lineskiplimit=0.9em\lineskip=0.7em plus 0.1em}  %
\usepackage{semantic}

\definecolor{dHilite}{rgb}{0.9, 0.9, 0.6}
\definecolor{dRed}{rgb}{0.45, 0.0, 0.0}
\definecolor{dGreen}{rgb}{0.0, 0.65, 0.0}
\definecolor{dDkGreen}{rgb}{0.0, 0.35, 0.0}
\definecolor{dBlue}{rgb}{0.0, 0.0, 0.65}
\definecolor{dPurple}{rgb}{0.65, 0.0, 0.65}
\definecolor{dDigPurple}{rgb}{0.5, 0.0, 0.5}
\definecolor{dFaint}{rgb}{0.7, 0.7, 0.7}
\definecolor{dGray}{rgb}{0.5, 0.5, 0.5}
\definecolor{dDark}{rgb}{0.2, 0.2, 0.2}
\definecolor{dAlmostBlack}{rgb}{0.1, 0.1, 0.1}

\makeatletter
\def\url@MGstyle{%
\def\UrlFont{\tiny\huge\ttfamily}%
\Url@do
}
\makeatother

\makeatletter
\def\url@vttstyle{%
  \@ifundefined{selectfont}{\def\UrlFont{\tt}}{\def\UrlFont{\normalfont\fontfamily{cmvtt}\selectfont}}}
\makeatother
\newcommand{\LoudLabel}[1]{\idempotentlabel{#1}%
\ifnum\OPTIONLoudLabels=1%
  \ifnum\OPTIONConf=1%
  \marginnote{\tiny\textvtt{#1}}%
  \else%
  \marginnote{\textvtt{#1}}%
  \fi%
\fi%
}

\newcommand{\idempotentlabel}[1]{%
    \ifcsname IDEMPFLAG#1\endcsname%
      \message{YYY ALREADY DEFINED: #1}
    \else%
      \message{YYZ NOT ALREADY DEFINED: #1}
      \expandafter\gdef\csname IDEMPFLAG#1\endcsname{d}%
      \label{#1}%
    \fi}

\ifnum\OPTIONLoudLabels=1%
\else%
\fi

\def\CompactJudgments{0}
\newcommand{\ctxoutsym}{\ifnum\CompactJudgments=1%
    \dashv%
  \else%
     \dashv%
  \fi}
\newcommand{\ctxout}[1]{\mathrel{\ctxoutsym}{#1}}

\newcommand{\J}{\mathcal{J}}

\newcommand{\emptyctx}{\cdot}

\newcommand{\matchor}{\ensuremath{\normalfont\,\texttt{|}\hspace{-5.35pt}\texttt{|}\,}}

\newcommand{\unitty}{\tyname{unit}}
\newcommand{\unitexp}{\text{\normalfont \tt()}}
\newcommand{\injexp}[2]{\keyword{inj}_{#1}\,{#2}}

\newcommand{\sumcaseexp}[5]{%
   \keyword{case}({#1},\; %
   \injexp{1}{#2}.\,{#3},\;
   \injexp{2}{#4}.\,{#5}%
   )%
   }
\newcommand{\sumcasedots}[1]{%
   \keyword{case}({#1},\;\cdots%
   )%
   }
\newcommand{\sumcaseexpv}[5]{%
   \keyword{case}({#1}, %
     \arrayenvl{%
       \injexp{1}{#2}.\,{#3}, \\%
       \injexp{2}{#4}.\,{#5}%
       )%
     }%
   }
\newcommand{\caseexp}[2]{\keyword{case}({#1},\;{#2})}
\let\case\sumcaseexp
\newcommand{\annoexp}[2]{\hspace*{-0.1ex}\texttt({#1} : {#2}\texttt)\hspace*{-0.1ex}}

\newcommand{\subtypingycolor}[1]{\textcolor{dDigPurple}{#1}}
\let\subtype\undefined
\newcommand{\subtype}{\mathrel{\normalfont\texttt{\subtypingycolor{<:}}}}  %

\newcommand{\proofheading}[1]{}  %

\newcommand{\AllSym}{\forall}
\newcommand{\xAll}[1]{\AllSym#1}
\newcommand{\All}[1]{\xAll{#1}.\:}

\newcommand{\subjudg}[4]{\ensuremath{{#1} \entails {#2} \subtype {#3} \ctxout{#4}}}

\newdimen\zzinstsymLTwidth
\newdimen\zzinstsymEQwidth
\newdimen\zzinstsymDiff

\newcommand{\chkcolor}{dBlue}
\newcommand{\syncolor}{dRed}
\newcommand{\appcolor}{dDkGreen}
\newcommand{\chk}{\mathrel{\mathcolor{\chkcolor}{\Leftarrow}}}
\newcommand{\uncoloredsyn}{{\Rightarrow}}
\newcommand{\syn}{\mathrel{\mathcolor{\syncolor}{\uncoloredsyn}}}
\newcommand{\appsep}{\;{\mathcolor{\appcolor}{\bullet}}\;}
\newcommand{\app}{\mathrel{\mathcolor{\appcolor}{{\uncoloredsyn}\hspace{-1.2ex}{\uncoloredsyn}}}}

\newcommand{\chkjudg}[4]{\ensuremath{{#1} \entails {#2} \chk {#3} \ctxout{#4}}}

\newcommand{\synjudg}[4]{\ensuremath{{#1} \entails {#2} \syn {#3} \ctxout{#4}}}

\newcommand{\declappjudg}[4]{\ensuremath{#1} \entails {#3} \appsep {#2}  \app {#4}}

\newcommand{\hypeq}[2]{{#1} = {#2}}

\newcommand{\extendssym}{\longrightarrow}
\newcommand{\extends}[2]{{#1} \extendssym {#2}}
\newcommand{\substextend}[2]{\extends{#1}{#2}}

\newcommand{\ahat}{\hat{\alpha}}
\newcommand{\bhat}{\hat{\beta}}

\newcommand{\rulename}[1]{\text{\normalfont\textsf{#1}}}

\newcommand{\Tyrulename}[1]{\ensuremath{\rulename{#1}}\xspace}

\newcommand{\Chkrulename}[1]{\Tyrulename{#1\ensuremath{\chk}}}
\newcommand{\Synrulename}[1]{\Tyrulename{#1\ensuremath{\syn}}}

\newcommand{\ChkIntrorulename}[1]{\Chkrulename{\ensuremath{#1}I}}

\newcommand{\SynElimrulename}[1]{\Synrulename{\ensuremath{#1}E}}

\newcommand{\Introrulename}[1]{\Tyrulename{\ensuremath{#1}I}}

\newcommand{\Elimrulename}[1]{\Tyrulename{\ensuremath{#1}E}}

\newcommand{\Var}{\Tyrulename{Var}}
\newcommand{\SynVar}{\Synrulename{Var}}
\newcommand{\Sub}{\Tyrulename{Sub}}
\newcommand{\ChkSub}{\Chkrulename{Sub}}
\newcommand{\Anno}{\Tyrulename{Anno}}
\newcommand{\SynAnno}{\Synrulename{Anno}}

\newcommand{\TypeEq}{\Tyrulename{TypeEq}}

\newcommand{\UnitIntro}{\Introrulename{\unitty}}
\newcommand{\ChkUnitIntro}{\ChkIntrorulename{\unitty}}

\newcommand{\ArrIntro}{\Introrulename{\arr}}
\newcommand{\ChkArrIntro}{\ChkIntrorulename{\arr}}

\newcommand{\ArrElim}{\Elimrulename{\arr}}
\newcommand{\SynArrElim}{\SynElimrulename{\arr}}

\newcommand{\LOCALCOPY}[1]{%
          \href{papers/#1}{\bf \textcolor{dGreen}{local copy}}}

\newcommand{\botty}{\bot}
\newcommand{\topty}{\top}

\newcommand{\Int}{\tyname{int}}
\newcommand{\Bool}{\tyname{bool}}

\newcommand{\sectty}{\mathbin{\land}}

\newcommand{\imp}{\supset}

\newcommand{\such}{\mathrel{{|}}}

\newrulecommand{SumElim}{$+$Elim}

\newcommand{\oderskyinhsym}{\lor}
\newcommand{\oderskysynsym}{\land}

\newcommand{\oderskyinh}[1]{{}^{\oderskyinhsym}{#1}}
\newcommand{\oderskysyn}[1]{{}_{\oderskysynsym}{#1}}

\newcommand{\Piercechk}{\stackrel{\leftarrow}{\in}}
\newcommand{\Piercesyn}{\stackrel{\rightarrow}{\in}}

\newcommand{\DaviesThesischk}{\stackrel{\Leftarrow}{\in}}
\newcommand{\DaviesThesissyn}{\stackrel{\Rightarrow}{\in}}

\newcommand{\annoextsym}{\sqsupseteq}
\newcommand{\annoext}{\mathrel{\annoextsym}}

\mathlig{|-}{\entails}
\mathlig{-|}{\ctxoutsym}
\mathlig{->}{\arr}
\mathlig{<<}{\langle}
\mathlig{>>}{\rangle}
\mathlig{<=}{\chk}
\mathlig{=>}{\syn}
\mathlig{=>>}{\app}

\newcommand{\From}{\Leftarrow}
\newcommand{\Rule}[2]{\Infer{}{#1}{#2}}
\newcommand{\CHECK}[3]{{#1} \vdash {#2} \chk {#3}}
\newcommand{\synth}[3]{{#1} \vdash {#2} \syn {#3}}
\let\inj\injexp
\newcommand{\Match}[2]{\mathsf{match}\;{#1}\;\mathsf{of}\;[ #2 ]}
\newcommand{\arm}[2]{{#1} \to {#2}}
\newcommand{\downshift}[1]{{\downarrow}\hspace*{0.45ex}{#1}}
\newcommand{\upshift}[1]{{\uparrow}{#1}}
\newcommand{\thunk}[1]{\left\{{#1}\right\}}
\newcommand{\return}[1]{\mathsf{return}\;#1}
\newcommand{\fun}[1]{\lambda{#1}.\,}
\newcommand{\CHECKn}[3]{{#1} \rhd {#2} \From {#3}}
\newcommand{\CHECKp}[2]{{#1} \leadsto {#2}}
\newcommand{\spine}[4]{{#1} \vdash {#2} : {#3} \gg {#4}}
\newcommand{\unit}{\unitexp}

\newcommand{\tensor}{\otimes}
\newcommand{\lolli}{\multimap}
\newcommand{\letv}[2]{\mathsf{let}\,{#1}={#2}\,\mathsf{in}\,}
\newcommand{\letunit}[1]{\letv{\unit}{#1}}
\newcommand{\pair}[2]{\left\langle {#1}, {#2} \right\rangle}
\newcommand{\letpair}[3]{\letv{\pair{#1}{#2}}{#3}}

\declaretheoremstyle[
  bodyfont=\sl
]{mytheoremstyle}

\declaretheorem[style=mytheoremstyle]{theorem}

\declaretheorem[style=mytheoremstyle]{definition}

\newtheoremstyle{better}%
   {\topsep}%
   {\topsep}%
   {\upshape\slshape}%
   {}%
   {\bfseries}%
   {.}%
   {0.7em}%
   {}%
\theoremstyle{better}
\newtheorem*{lemma*}{Lemma}
\newtheorem*{definition*}{Definition}

\usepackage{booktabs} %

\usepackage[ruled]{algorithm2e} %

\SetAlFnt{\small}
\SetAlCapFnt{\small}
\SetAlCapNameFnt{\small}
\SetAlCapHSkip{0pt}
\IncMargin{-\parindent}

\ifnum\OPTIONArxiv=1
  \setcopyright{none}
\else
  \acmJournal{CSUR}
\copyrightyear{2019}
\setcopyright{acmlicensed}


\fi

\begin{document}
\title{%
  Bidirectional Typing%
}

\author[Jana Dunfield]{Jana Dunfield}
\orcid{0000-0002-3718-3395}
\affiliation{%
  \institution{Queen's University}
  \streetaddress{School of Computing, Goodwin Hall 557}
  \city{Kingston, ON}
  \postcode{K7L 3N6}
  \country{Canada}
}
\email{jd169@queensu.ca}

\author[Neel Krishnaswami]{Neel Krishnaswami}
\orcid{0000-0003-2838-5865}
\affiliation{%
  \institution{University of Cambridge}
  \streetaddress{Computer Laboratory, William Gates Building}
  \city{Cambridge}
  \postcode{CB3 0FD}
  \country{United Kingdom}}
\email{nk480@cl.cam.ac.uk}

\begin{abstract}
  Bidirectional typing combines two modes of typing: type checking, which checks
  that a program satisfies a known type,
  and type synthesis, which determines a type from the program.
  Using checking enables bidirectional typing to support features for which
  inference is undecidable;
  using synthesis enables bidirectional typing to avoid the large annotation burden
  of explicitly typed languages.
  In addition, bidirectional typing improves error locality. %
  We highlight the design principles that underlie bidirectional type systems,
  survey the development of bidirectional typing
  from the prehistoric period before Pierce and Turner's local type inference
  to the present day,
  and provide guidance for future investigations.   %
\end{abstract}

\maketitle

\setcounter{footnote}{0}

\section{Introduction}
\label{sec:intro}

Type systems serve many purposes.
They allow programming languages to reject nonsensical programs.
They allow programmers to express their intent, and to use a type checker
to verify that their programs are consistent with that intent.
Type systems can also be used to automatically insert implicit operations,
and even to guide program synthesis.

Automated deduction and logic programming give us a useful lens
through which to view type systems: \emph{modes} \citep{Warren77}.
When we implement a typing judgment, say $\Gamma |- e : A$,
is each of the meta-variables ($\Gamma$, $e$, $A$)
an input, or an output?
If the typing context $\Gamma$, the term $e$ and the type $A$ are inputs,
we are implementing type checking.
If the type $A$ is an output, we are implementing type inference.
(If only $e$ is input, we are implementing \emph{typing inference}
\citep{Jim95:WhatArePT,Wells02:EssenceOfPT};
if $e$ is output, we are implementing program synthesis.)
The status of each meta-variable---input or output---is its \emph{mode}.

As a general rule, outputs make life more difficult.  In complexity theory,
it is often relatively easy to check that a given solution is valid, but finding
(synthesizing) a solution may be complex or even undecidable.
This general rule holds for type systems:
synthesizing types may be convenient for the programmer,
but computationally intractable.

To go beyond the specific feature set of traditional Damas--Milner typing,
it might seem necessary to abandon synthesis%
\footnote{We choose to say \emph{synthesis} instead of \emph{inference}.
This is less consistent with one established usage, ``type inference'', 
but more consistent with another, ``program synthesis''.}%
.  %
Instead, however, we can \emph{combine}
synthesis with checking.  In this approach, \emph{bidirectional typing},
language designers are not forced to choose between a rich set of typing features
and a reasonable volume of type annotations: implementations of
bidirectional type systems alternate between treating the type as input,
and treating the type as output.

The practice of bidirectional typing has, at times, exceeded its foundations:
the first commonly cited paper on bidirectional typing appeared in 1997
but mentioned that the idea was known as ``folklore'' (see \Sectionref{sec:history}).
Over the next few years, several bidirectional systems appeared,
but the principles used to design them were not always made clear.
Some work did present underlying design principles---but within the
setting of some particular type system with other features of interest,
rather than focusing on bidirectional typing as such.
For example, \citet{Dunfield04:Tridirectional} gave a broadly applicable
design recipe for bidirectional typing, but their focus was on an idea---typing rules
that decompose evaluation contexts---that has been applied more narrowly.

Our survey has two main goals:

\begin{enumerate}
\item to collect and clearly explain the design principles of bidirectional typing,
  to the extent they have been discovered; and

\item to provide an organized summary of past research related to bidirectional typing.
\end{enumerate}

We begin by describing a tiny bidirectional type system (\Sectionref{sec:stlc}).
\Sectionref{sec:elements} presents some design criteria for bidirectional type systems.
\Sectionref{sec:recipe} describes a modified version of
the recipe of \citet{Dunfield04:Tridirectional},
and relates it to our design criteria.
\Sectionref{sec:polymorphism} discusses work on combining (implicit) polymorphism
with bidirectional typing,
and \Sectionref{sec:variations} surveys other variations on bidirectional typing.
Sections 
\ref{sec:proof-theory}--\ref{sec:focusing}
give an account of connections between bidirectional typing and topics such as
proof theory,
focusing and polarity, and 
call-by-push-value. 
Section \ref{sec:applications} cites other work that uses bidirectional typing.
We conclude with historical notes (\Sectionref{sec:history})
and
a summary of notation (\Sectionref{sec:notation}).

\section{Bidirectional Simply Typed Lambda Calculus}
\label{sec:stlc}

To develop our first bidirectional type system,
we start with a (non-bidirectional) simply typed lambda calculus (STLC).
This STLC is not the smallest possible calculus:
we include a term $\unitexp$ of type $\unitty$,
to elucidate the process of bidirectionalization.
We also include a gratuitous ``type equality'' rule.  %

\begin{figure}[t]
  \centering
  
  \begin{grammar}
    Expressions
    & $e$
    & \bnfas
    &
        $
        x
        \bnfalt
        \Lam{x} e
        \bnfalt
        e\,e
        \bnfalt
        \unitexp
        $
    \\
    Types
    & $A, B, C$
    & \bnfas
    &
       $
       \unitty
       \bnfalt
       A -> A
       $
    \\
    Typing contexts
    & $\Gamma$
    & \bnfas
    &
       $
       \emptyctx
       \bnfalt
       \Gamma, x : A
       $
  \vspace*{-1.0ex}
  \end{grammar}

  \hspace*{-0.030\linewidth}\begin{minipage}[b]{0.375\linewidth}
      \judgbox{\arrayenvl{
        \Gamma |- e : A
      }}%
         {Under context $\Gamma$,
           \\
           expression $e$ has type $A$
         }
      \begin{mathpar}
        \Infer{\Var}
            {(x : A) \in \Gamma}
            {\Gamma |- x : A}
        \\
        \Infer{\TypeEq}
            {
              \Gamma |- e : A
              \\
              A = B
            }
            {\Gamma |- e : B}
        \\
        \Infer{\Anno}
            {
              \Gamma |- e : A
            }
            {\Gamma |- \annoexp{e}{A} : A}
        \vspace*{-2ex}
        \\
        \Infer{\UnitIntro}
            {
            }
            {\Gamma |- \unitexp : \unitty}
        \\
        \Infer{\ArrIntro}
            {
              \Gamma, x : A_1 |- e : A_2
            }
            {
              \Gamma |- (\Lam{x} e) : A_1 -> A_2
            }
        \and
        \Infer{\ArrElim}
            {
              \Gamma |- e_1 : A -> B
              \\
              \Gamma |- e_2 : A
            }
            {\Gamma |- e_1\,e_2 : B}
      \end{mathpar}
  \end{minipage}
  ~\!
  \begin{minipage}[b]{0.56\linewidth}
      \judgbox{
        \Gamma |- e <= A
        \\
        \Gamma |- e => A
      }%
      {  ~\\[0.3ex]
        Under $\Gamma$, expression $e$ checks against type $A$
           \\[0.4ex]
           Under $\Gamma$, expression $e$ synthesizes type $A$
      }
      \vspace*{1.0ex}
      \begin{mathpar}
        \Infer{\SynVar}
            {(x : A) \in \Gamma}
            {\Gamma |- x => A}
        \\
        \Infer{\ChkSub}
            {
              \Gamma |- e => A
              \\
              A = B
            }
            {\Gamma |- e <= B}
        \\
        \Infer{\SynAnno}
            {
              \Gamma |- e <= A
            }
            {\Gamma |- \annoexp{e}{A} => A}
        \vspace*{-2ex}
        \\
        \Infer{\ChkUnitIntro}
            {
            }
            {\Gamma |- \unitexp <= \unitty}
        \\
        \Infer{\ChkArrIntro}
            {
              \Gamma, x : A_1 |- e <= A_2
            }
            {
              \Gamma |- (\Lam{x} e) <= A_1 -> A_2
            }
        \and
        \Infer{\SynArrElim}
            {
              \Gamma |- e_1 => A -> B
              \\
              \Gamma |- e_2 <= A
            }
            {\Gamma |- e_1\,e_2 => B}
      \end{mathpar}
  \end{minipage}

  \caption{A simply typed $\lambda$-calculus ($:$ judgment)
    and a bidirectional version ($=>$ and $<=$ judgments)}
  \label{fig:stlc}
\end{figure}

Our non-bidirectional STLC (\Figureref{fig:stlc}, left side)
has six rules deriving the judgment $\Gamma |- e : A$:
a variable rule,
a type equality rule,
a rule for type annotations,
an introduction rule for $\unitty$,
an introduction rule for $->$
and an elimination rule for $->$.

These rules are standard except for the type equality rule \TypeEq,
which says that if $e$ has type $A$
and $A$ equals $B$, then $e$ has type $B$.
If this rule does not disturb you, you may skip to the next paragraph.
If we had polymorphism, we could motivate this rule by arguing
that $\All{\alpha} \alpha -> \alpha$ should be considered equal to
$\All{\beta} \beta -> \beta$.   %
Our language of types, however, has only $\unitty$ and $->$
and so admits no nontrivial equalities.
The role of \TypeEq is retrospective:
it is the type assignment version of a necessary bidirectional rule,
allowing us to tell a more uniform story of making type assignment bidirectional.
For now, we only note that the type equality rule is sound:
for example, if we have derived $\Gamma |- e : \unitty -> \unitty$
then $\Gamma |- e : \unitty -> \unitty$, which was already derived.

Given these six STLC typing rules,
we produce each bidirectional rule in turn
(treating the type equality rule last).
Some of our design choices will become clear only in the light of
the ``recipe'' in \Sectionref{sec:recipe}, but the recipe would not be clear
without seeing a bidirectional type system first.

\begin{enumerate}
\item
  The variable rule \Var has no typing premise,
  so our only decision is whether the conclusion should
  synthesize $A$ or check against $A$.
  The information that $x$ has type $A$ is in $\Gamma$,
  so we synthesize $A$.
  A checking rule would require that the type be known
  \emph{from the enclosing term},
  a very strong restriction:
  even $f\;x$ would require a type annotation.

\item
  From the annotation rule \Anno
  we produce \SynAnno, which synthesizes its conclusion:
  we have the type $A$ in $\annoexp{e}{A}$,
  so we do not need $A$ to be given as input.
  In the premise, we check $e$ against $A$;
  synthesizing $A$ would prevent the rule from typing a non-synthesizing $e$,
  which would defeat the purpose of the annotation.
  
\item
  Unit introduction
  \UnitIntro
  \emph{checks}.
  At this point in the paper,
  we prioritize internal consistency:
  checking $\unitexp$ is consistent with
  the introduction rule for $->$ (discussed next),
  and with the introduction rule for products. %

\item
  Arrow introduction
  \ArrIntro
  checks.
  This decision is better motivated:
  to synthesize $A_1 -> A_2$ for $\lam{x} e$
  we would have to synthesize a type for the body $e$.
  That raises two issues.
  First, by requiring that the body synthesize,
  we would need a second rule to handle $\lambda$s
  whose body checks but does not synthesize.
  Second, if we are synthesizing $A_1 -> A_2$
  that means \emph{we don't know $A_1$ yet},
  which prevents us from building the context $\Gamma, x : A_1$.
  (Placeholder mechanisms, which allow building $\Gamma, x : \ahat$
  and ``solving'' $\ahat$ later, are described in \Sectionref{sec:polymorphism}.)

  Since the conclusion is checking,
  we know $A_2$, so we might as well check in the premise.

\item
  For arrow elimination
  \ArrElim,
  the \emph{principal judgment} is the premise
  $\Gamma |- e_1 : A -> B$,
  because that premise contains the connective being eliminated.
  We make that judgment synthesize;
  this choice is the one suggested by our ``recipe'',
  and happens to work nicely:
  If $e_1$ synthesizes $A -> B$,
  we have $A$ and can check the argument
  (so the rule will work even when $e_2$ cannot synthesize),
  and we have $B$ so we can synthesize $B$ in the conclusion.
  (It is possible to have the premise typing $e_1$ be a checking judgment.
  In that case, the argument $e_2$ \emph{must} synthesize,
  because we need to know what $A$ is to check $e_1$ against $A -> B$.
  Similarly, the conclusion must be checking, because we need to know $B$;
  see \Sectionref{sec:arguments-first}.)

\item
  Finally, we come to the type equality rule \TypeEq.
  Where the type assignment premise
  $\Gamma |- e : A$ and
  conclusion $\Gamma |- e : B$ are identical (since $A$ is exactly equal to $B$),
  the duplication of these identical premises enables us
  to give them different directions in the bidirectional system.
  Either
  (1) the conclusion should synthesize (and the premise check),
  or
  (2) the conclusion should check (and the premise synthesize).

  Option (1) cannot be implemented:
  If the conclusion synthesizes $B$,
  that means $B$ is \emph{not} an input;
  we don't know $B$, which means we also don't know $A$
  for checking.

  Option (2) works:
  If we want to check $e$ against a known $B$ in the conclusion,
  and $e$ synthesizes a type $A$, we verify that $A = B$.
\end{enumerate}

Neither \ChkSub nor \SynAnno is tied to an operational feature
(as, for instance, \SynArrElim is tied to functions);
\SynAnno is tied to a syntactic form,
but (supposing a type erasure semantics)
not to any operational feature.
Moreover, \ChkSub and \SynAnno have a certain symmetry:
\ChkSub moves from a checking conclusion
to a synthesizing premise,
while \SynAnno moves from a synthesizing conclusion
to a checking premise.

\section{Elements of Bidirectional Typing}
\label{sec:elements}

From a rules-crafting perspective, bidirectionality adds a degree of design freedom to every
judgment (premises and conclusion) in a rule:
should a particular premise or conclusion synthesize a type, or check against a known type?
Covering all possibilities with rules that have every combination of synthesis and checking
judgments would lead to an excessive number of rules.
For example, the following eight rules are superficially valid bidirectional
versions of the standard $->$-elimination rule.
\newcommand{\RenderRule}[4]{%
  \Infer{#1}
      {
        \arrayenvl{
          \Gamma |- e_1 #3 A -> B
          \\
          \Gamma |- e_2 #4 A
        }
      }
      {\Gamma |- e_1\,e_2 #2 B}
}%
\begin{mathpar}
  \RenderRule{}{=>}{=>}{=>}
  \and
  \RenderRule{}{=>}{=>}{<=}
  \and
  \RenderRule{}{=>}{<=}{=>}
  \and
  \RenderRule{}{=>}{<=}{<=}
  \\
  \RenderRule{}{<=}{=>}{=>}
  \and
  \RenderRule{}{<=}{=>}{<=}
  \and
  \RenderRule{}{<=}{<=}{=>}
  \and
  \RenderRule{}{<=}{<=}{<=}
\end{mathpar}
What criteria should guide the designer in crafting a manageable set of rules
with good practical (and theoretical) properties?

\subsection{First Criterion: Mode-correctness}

The first criterion comes from logic programming \citep{Warren77}.
We want to avoid having to guess types: in an ideal world,
whenever we synthesize a type, the type should come from known information---rather than,
say, enumerating all possible types.
A rule is \emph{mode-correct} if there is a strategy for recursively
deriving the premises such that two conditions hold:

\begin{enumerate}[(1)]
\item The premises are mode-correct: 
  for each premise, every input meta-variable is known (from the inputs to 
  the rule's conclusion and the outputs of earlier premises).
\item
  The conclusion is mode-correct:
  if all premises have been derived, the outputs of the conclusion are known.
\end{enumerate}

Our last rule, in which every judgment is checking ($<=$),
is \emph{not} mode-correct:
In the first premise $\Gamma |- e_1 <= A -> B$,
the context $\Gamma$ and term $e_1$ are known from the inputs $\Gamma$ and $e_1\,e_2$
in the conclusion $\Gamma |- e_1\,e_2 <= B$.
However, the type $A -> B$ cannot be constructed, because $A$ is not known.
For the same reason, the second premise $\Gamma |- e_2 <= A$
is not mode-correct.
(The conclusion is mode-correct because all the meta-variables are inputs.)

Only four of the above eight rules are mode-correct:
\begin{mathpar}
  \RenderRule{}{=>}{=>}{=>}
  \and
  \RenderRule{}{=>}{=>}{<=}
  \\
  \RenderRule{}{<=}{=>}{<=}
   \and
   \RenderRule{}{<=}{<=}{=>}
\end{mathpar}
The mode-correctness of the fourth rule seems questionable:
if we insist on recursively deriving the first premise
$\Gamma |- e_1 <= A -> B$
before the second premise
$\Gamma |- e_2 => A$,
it is not mode-correct,
but if we swap the premises and view the rule as
\begin{mathpar}
  \Infer{}
      {
        \Gamma |- e_2 => A
        \\
        \Gamma |- e_1 <= A -> B
      }
      {\Gamma |- e_1\,e_2 <= B}
\end{mathpar}
it is mode-correct: the subterm $e_2$ synthesizes $A$,
and $B$ is an input in the conclusion,
so $A -> B$ is known and $e_1$ can be checked against it.

This seems disturbing: the order of premises should not affect the meaning of the rule,
in the sense that a rule determines a set of possible derivations.
But mode-correctness is not about meaning in that sense; rather, it is about a particular strategy
for applying the rule.  For any one strategy, we can say whether the strategy is mode-correct;
then we can say that a rule is mode-correct if there exists some strategy that is mode-correct.
The set of strategies is the set of permutations of the premises.  So the rule above has
two strategies, one for each of the two permutations of its premises; since one of the strategies
is mode-correct, the rule is mode-correct.

A bidirectional type system that is \emph{not} mode-correct cannot be directly
implemented, defeating the goal that the derivability of a typing judgment
should be %
decidable.
Thus, mode-correctness is necessary.
However, mode-correctness alone does not always lead to a practical algorithm:
if more than one rule is potentially applicable, a direct implementation
requires backtracking.
When the inputs (the context, the term, and---if in the checking direction---the type)
match the conclusion of only one rule, the system is syntax-directed:
we can ``read off'' an implementation from the rules.

\subsection{Second Criterion: Completeness (Annotatability)}

The empty set of rules satisfies the first criterion: %
every rule is mode-correct, because there are no rules.  %

Our second criterion rejects the empty system.
A bidirectional system is \emph{complete with respect to a type assignment system}
if every use of a rule in the type assignment system can be ``matched'' by some rule in the
bidirectional system.  This matching is approximate, because applying the bidirectional rule
might require that we change the term---generally, by adding type annotations
(sometimes called ascriptions).

For example, forgetting to include an $->$-elimination rule in
a bidirectional type system would make the system incomplete:
it would reject all function applications.
In general, completeness is easy to achieve, provided we begin with a
type assignment system and ``bidirectionalize'' each rule.

Because the move from a type assignment derivation to a bidirectional derivation
may require adding annotations, the related theorem is sometimes called annotatability
instead of completeness.
A separate criterion considers the quantity and quality of the required annotations
(\Sectionref{sec:crit:annotation-character}).

Requiring that every type connective have at least one introduction rule and at least one elimination rule
would be too strict: the empty type $\botty$ should not have an introduction rule,
and the top type $\topty$ and the unit type do not require elimination rules.
(We might choose to include an elimination rule for $\topty$, but our criteria for bidirectional systems
should not force this choice.)

\subsection{Third Criterion: Size}

Our third criterion refers to the number of typing rules.
This is not always a reliable measure; by playing games with notation,
one can inflate or deflate the number of typing rules.
But the measure can be used effectively when comparing two systems
in an inclusion relationship:
a system of two rules R1 and R2 is clearly smaller than a system that
has R1, R2, and a third rule R3.
Smaller systems tend to be easier to work with:
for example,
in a meta-theoretic proof that considers cases of the rule concluding a given derivation,
each new rule leads to an additional proof case.

\subsection{Fourth Criterion: Annotation Character}
\label{sec:crit:annotation-character}

A bidirectional system that required an annotation on \emph{every} subterm
would satisfy completeness.
To rule out such a system, we need another criterion.
We call it \emph{annotation character},
an umbrella term for attributes that are sometimes in conflict:

\begin{enumerate}[(i)]
\item %
  Annotations should be \emph{lightweight}: they should constitute a small
  portion of the program text.
\item %
  Annotations should be \emph{predictable}: programmers should be able
  to easily determine whether a subterm needs an annotation.
  That is, there should be a clear \emph{annotation discipline}.
\item %
  Annotations should be \emph{stable}:
  a small change to a program should have a small effect on
  annotation requirements.

\item  %
  Annotations should be \emph{legible}: the form of annotation should
  be easy to understand.
\end{enumerate}

Attribute (i) is the easiest to measure, but that doesn't make it the most important.

Attribute (ii) is harder to measure, because it depends on the definition of ``easily''
(alternatively, on the definition of ``clear'').
In the absence of empirical studies comparing bidirectional type systems with
different annotation disciplines, we can form some hypotheses:

\begin{enumerate}[(1)]
\item A discipline that needs only \emph{local} information
  is preferable to one that needs global information.
  That is, we want to know whether a subterm needs annotation
  from the ``neighbourhood'' of that subterm, not by looking at the
  whole program.
  The Pfenning recipe (\Sectionref{sec:recipe}) leads to an annotation discipline in which
  the syntactic forms of the subterm (\eg that it is a pair)
  and the subterm's immediate context (its parent in the syntax tree)
  suffice to determine whether the subterm needs an annotation.
  (The subterm alone is not enough:
  a subterm that cannot synthesize does not require an annotation
  if it is being checked, and whether the subterm is checked depends
  on its position in the program.)

  In an annotation discipline that needs only local information,
  a change to one part of a program cannot affect the need for annotations
  in distant parts of the program.
  Hence, such a discipline is \emph{stable} (attribute (iii)).

\item A discipline that requires all \emph{non-obvious} type annotations,
  and no \emph{obvious} type annotations, is preferable.
  Unfortunately,
  it is not easy to agree on which type annotations are obvious.

\item A discipline that needs obvious type annotations in certain situations is acceptable,
  if those situations are rare.
  For example, we might tolerate annotations on every \textkw{while} loop in SML
  because SML programmers rarely use \textkw{while}.
\end{enumerate}

These hypotheses can be found, in a different form, in earlier work.
The term \emph{local type inference} \citep{Pierce00} implies that
bidirectional typing should focus on (or even be restricted to) local information,
suggesting that annotation disciplines should not need global information.
Hypotheses (2) and (3) correspond to this explanation, from the same paper:

\begin{quote}
  The job of a partial type inference algorithm should be to eliminate especially
  those type annotations that are both \emph{common} and \emph{silly}---i.e., those
  that can be neither justified on the basis of their value as checked documentation
  nor ignored because they are rare. \citep[\S1.1; emphasis in original]{Pierce00}
\end{quote}

\section{A Bidirectional Recipe}
\label{sec:recipe}

Fortunately, we have a methodical approach that produces bidirectional systems
that satisfy many of the above criteria.

We call this approach\footnote{%
Our presentation of the recipe is intended as
a more detailed explanation of the original,
with one exception.
The exception is that, instead of (implicitly) advocating
that a \keyword{case} expression (sum elimination) have a single rule with a checking conclusion,
our version of the recipe allows for \emph{two} rules: one checking, one synthesizing.%
} %
the \emph{Pfenning recipe} \citep{Dunfield04:Tridirectional}.  
It yields a set of bidirectional typing rules that is
\emph{small} (our second criterion)
and whose annotation discipline is
moderately \emph{lightweight} and highly \emph{predictable}
(attributes of our fourth criterion).
Some disadvantages of the recipe---particularly in terms of a lightweight annotation discipline---%
can be mitigated, in exchange for a larger set of rules.
Thus, the foremost virtue of the recipe is that it gives a \emph{starting point} for a practical system:
it tells you what the ground floor of the building (type system) should look like,
allowing a taller building if desired.
Even if the recipe is not followed completely, some of its steps are useful in designing
bidirectional systems; we use this opportunity to explain those steps.
Another virtue of the recipe is that it guarantees a subformula property,
which can be of practical as well as theoretical interest;
see \Sectionref{sec:subformula-property}.

Nonetheless, this recipe is not the only way to design a bidirectional type system.
This fact is demonstrated by the wide variety of systems that do not exactly follow
the recipe, and by the existence of systems that diverge radically from it (\Sectionref{sec:backwards}).

\subsection{Introduction and Elimination Rules}
\label{sec:recipe-intro-elim}

This part of the recipe pertains to each rule that is an introduction rule
(introducing a type connective that occurs in the conclusion)
or an elimination rule (eliminating a type connective that occurs in a premise).

\paragraph{Step 1: Find the principal judgment} 
The \emph{principal connective} of an introduction (resp.\ elimination) rule
is the connective that is being introduced (resp.\ eliminated).
The \emph{principal judgment} of a rule is the judgment containing the principal connective;
in an introduction rule, the principal judgment is usually%
\footnote{In an introduction rule in which the introduced connective
is only available in a lexically scoped subterm,
the principal connective is added to the context in a premise.
In such a rule, the premise with the extended context---%
not the conclusion---is the principal judgment.
An example is a rule typing a product introduction for a ``let-pair'' construct:
the (highlighted) third premise is the principal judgment,
because it contains the introduced connective $\times$.
\[
\Infer{}
    {
      \Gamma |- e_1 : A_1
      \\
      \Gamma |- e_2 : A_2
      \\
      \fighi{\Gamma, x : A_1 \times A_2 |- e : B}
    }
    {\Gamma |- \textkw{let}\,x = \langle e_1, e_2 \rangle\;\textkw{in}\,e : B}
\]
}
the conclusion, and in an elimination rule, the principal judgment is (as far as we know)
always the first premise.

\paragraph{Step 2: Bidirectionalize the principal judgment}
This is the magic ingredient!
If the rule is an \emph{introduction} rule, make the principal judgment \emph{checking}.
If the rule is an \emph{elimination} rule, make the principal judgment \emph{synthesizing}.

\paragraph{Step 3: Bidirectionalize the other judgments}
The direction of the principal judgment provides guidance for the other directions.
The first question is, in what order should we bidirectionalize the other judgments?
If the principal judgment is the conclusion (often true for introduction rules),
the only judgments left are the premises,
but if the principal judgment is a premise (probably the first premise),
we have a choice.
The better choice seems to be to bidirectionalize the premises,
then the conclusion: this maximizes the chance of having enough information
to synthesize the type of the conclusion.

The second question is, which directions should we choose?
Here we are guided by our criteria of annotatability and annotation character:
the bidirectional system should be complete
with respect to ``ground truth'' (roughly, the given type assignment system)
without needing too many annotations.

Therefore, we should \emph{utilize known information}:
If we already know the type of a judgment---%
the judgment should be checking.
Thus, if the conclusion checks against a type $A$
and a premise also has type $A$,
the premise should be checking.
Similarly, if an earlier premise synthesizes $B$
then a later premise having type $B$ should be checking.
Choosing to synthesize would ignore known information,
restricting the system's power for no reason.

\paragraph{Step 3, continued}

Unfortunately, what should count as ``ground truth'' is not always clear.
It may not be exactly the set of type assignment rules.
In our experience, most type assignment rules can be taken as ground truth
and transformed by the recipe with satisfactory results,
but certain type assignment rules must be viewed more critically.
Consider a standard sum-elimination rule,
on a pattern-matching term $\sumcaseexp{e}{x_1}{e_1}{x_2}{e_2}$
where $x_1$ is bound in $e_1$ and $x_2$ is bound in $e_2$:
\[
  \Infer{\SumElim}
      {
        \Gamma |- e : (A_1 + A_2)
        \\
        \arrayenvbl{
          \Gamma, x_1 : A_1 |- e_1 : B
          \\
          \Gamma, x_2 : A_2 |- e_2 : B
        }
      }
      {
        \Gamma |- \sumcaseexp{e}{x_1}{e_1}{x_2}{e_2} : B
      }
\]
If we erase the terms from this rule and rewrite $+$ as $\lor$, we get a logical or-elimination rule:
\[
  \Infer{$\lor$Elim}
      {
        \Gamma |- (A_1 \lor A_2)
        \\
        \arrayenvbl{
          \Gamma, A_1 |- B
          \\
          \Gamma, A_2 |- B
        }
      }
      {
        \Gamma |- B
      }
\]
This rule is an \emph{instance} of a fundamental logical principle: reasoning by cases.
In a mathematical proof,
reasoning by cases works the same regardless of the goal we want to prove:
whether we want to conclude
``formula $B$ is true''
or ``kind $\kappa$ is well-formed''
or ``real number $x$ is negative''
or ``machine $H$ halts'',
our proof can consider the two cases ($A_1$ is true; $A_2$ is true)
of the disjunctive formula $A_1 \lor A_2$.
So the ground principle from which $\lor$Elim is instantiated
allows for a conclusion $\J$ that has \emph{any} form:
\[
  \Infer{$\lor$Elim-general}
      {
        \Gamma |- (A_1 \lor A_2)
        \\
        \arrayenvbl{
          \Gamma, A_1 |- \J
          \\
          \Gamma, A_2 |- \J
        }
      }
      {
        \Gamma |- \J
      }
\]
Instantiating $\J$ to ``formula $B$ is true'' results in $\lor$Elim,
while instantiating $\J$ to ``machine $H$ halts'' would result in 
\[
  \Infer{$\lor$Elim-halting-goal}
      {
        \Gamma |- (A_1 \lor A_2)
        \\
        \arrayenvbl{
          \Gamma, A_1 |- H\text{~halts}
          \\
          \Gamma, A_2 |- H\text{~halts}
        }
      }
      {
        \Gamma |- H\text{~halts}
      }
\]
If we consider $\lor$Elim-general to be the basis of \SumElim,
we see that $\lor$Elim-general should give rise to \emph{two} bidirectional rules,
because a bidirectional system has two judgment forms.
\[
  \Infer{}%
      {
        \Gamma |- e => (A_1 + A_2)
        \\
        \arrayenvbl{
          \Gamma, x_1 : A_1 |- e_1 <= B
          \\
          \Gamma, x_2 : A_2 |- e_2 <= B
        }
      }
      {
        \Gamma |- \sumcaseexp{e}{x_1}{e_1}{x_2}{e_2} <= B
      }
  ~~~~~~
  \Infer{}%
      {
        \Gamma |- e => (A_1 + A_2)
        \\
        \arrayenvbl{
          \Gamma, x_1 : A_1 |- e_1 => B
          \\
          \Gamma, x_2 : A_2 |- e_2 => B
        }
      }
      {
        \Gamma |- \sumcaseexp{e}{x_1}{e_1}{x_2}{e_2} => B
      }
\]
The original recipe \citep{Dunfield04:Tridirectional} resulted in only one
typing rule for \textkw{case}, the checking rule; if you can only have one rule,
the checking rule is more general and nicely matches the checking premise of $->$Intro.
Note that whether we have one rule or two, we are still following Steps 1 and 2
of the recipe: in both versions of the rule, the principal judgment synthesizes.
The complication is that making the principal judgment synthesize does not
determine the directions of the conclusion and the other premises.

Similar considerations arise in typing a let-expression, with one additional
wrinkle: a let-expression is not an elimination form.
Thus, we seem to have no guidance at all, beyond mode-correctness.
Whether the conclusion checks or synthesizes, we do not yet know the type $A$
of the let-bound expression, so the first premise must synthesize
(similar to a \keyword{case} expression):
\[
  \Infer{}
     {
       \Gamma |- e => A
       \\
       \Gamma, x : A |- e' : B
     }
     {\Gamma |- \letv{x}{e}{e'} : B}
\]
Now we have the same choice as in case elimination:
if a let-expression represents a general reasoning principle,
then we may want two rules, one where the second premise and conclusion
are checking, and one where they synthesize.
If we prioritize a small number of rules, the checking rule is more general
than the synthesis rule alone.
(It is actually possible to design a system where $e$ is checked;
we explore that territory in \Sectionref{sec:backwards}.)

\citet{Dunfield04:Tridirectional} claimed that annotations were needed only on redexes,
but that claim was false.
The claim holds for elimination forms that do not bind variables, %
but fails with elimination forms that \emph{do} bind variables, %
like \keyword{case} expressions.
If we, instead, have both rules for \keyword{case} expressions,
using a variable-binding elimination to derive the principal judgment of
a variable-binding elimination does not incur an annotation
(so, unlike in the original recipe,
$\sumcasedots{\sumcasedots{x}}$
can be typed if all arms of the inner \keyword{case} can synthesize).
However, some combinations of binding and non-binding elimination
do incur annotations despite having no redexes.\footnote{%
Admittedly, the truth of the claim depends on the meaning of ``redex'';
if we include various commuting conversions in our notion of reduction,
the claim becomes accurate.
}
We discuss the details in \Sectionref{sec:recipe-annotation-character}.

\subsection{Annotation}

Assume we have, in the given type assignment system, the following annotation rule.
(If we had no concern for ease of implementation, we might not need such a rule.)
\[
  \Infer{}
     {\Gamma |- e : A}
     {\Gamma |- \annoexp{e}{A} : A}
\]
This is not an introduction or elimination rule, so it has no principal connective
and thus no principal judgment.
However, the conclusion feels closer to being the principal judgment,
because---while the type $A$ is not tied to any particular connective---the type
must match the type appearing in the term $\annoexp{e}{A}$.
In contrast, the premise $\Gamma |- e : A$ imposes no constraints at all.

Thus, we will start by bidirectionalizing the conclusion.
Again, the rule is neither an introduction nor an elimination so
we cannot pick a direction based on that.
Instead, we follow the principle in Step 3 (above):
we should try to use all available knowledge.
We know the type $A$ because it appears in the term $\annoexp{e}{A}$,
so synthesizing $A$ in the conclusion utilizes this knowledge.

Now we turn to the premise, $\Gamma |- e : A$.
Since $A$ is known we should utilize it
by checking $e$ against $A$, resulting in the rule
\[
  \Infer{}
     {\Gamma |- e <= A}
     {\Gamma |- \annoexp{e}{A} => A}
\]
It doesn't really matter whether we start with the conclusion or the premise.
If we start with the premise, we notice that $A$ is known from $\annoexp{e}{A}$
and make the premise checking;
then we notice that $A$ is known in the conclusion.

\subsection{Variables}

A typing rule for variables is neither an introduction nor elimination rule.
Instead, it corresponds to a fundamental principle of deduction: the use of an assumption.  
Instead of interpreting the assumption $x : A$ to mean that $x$ has (is assigned) type $A$,
we interpret $x : A$ as $x => A$: we assume that $x$ synthesizes $A$, and so the variable rule
is synthesizing:
\[
  \Infer{\Var}
      {(x : A) \in \Gamma}
      {\Gamma |- x => A}
\]
It might be more clear for the form of assumptions in $\Gamma$
to be $x => A$ rather than $x : A$,
making clear that this rule is simply the use of an assumption,
but the form $x : A$ is standard.  

Since $\lambda$ is checked, its argument type is known and can
be added to $\Gamma$.  Assumptions $x <= A$ arise in
a reversed bidirectional system (\Sectionref{sec:backwards}).

\subsection{Change of Direction (Subsumption)}
\label{sec:changedir}

To see why this part of the recipe is needed, consider the following type
assignment derivation.
\[
  \Infer{}
      {}
      {x : A |- x : A}
\]
Our bidirectional \Var rule can synthesize a type for $x$,
but cannot derive a checking judgment.
So we cannot synthesize a type for $f$ applied to $x$,
even though both their types are available in the typing context:
\[
  \Infer{Syn$->$Elim}
      {
        \Infer{\Var}
            {(f : A -> B) \in (f : A -> B,~ x : A)}
            {f : A -> B,~ x : A |- f => A -> B}
        \\
        f : A -> B,~ x : A \not\entails x <= A
      }
      {f : A -> B,~ x : A \not\entails f\,x => B}
\]
We know, from the type of $f$, that $x$ needs to have type $A$.
The var rule can tell us that $x$ also synthesizes that type.
So we need a rule that verifies that a fact ($x$ synthesizes type $A$)
is consistent with a requirement ($x$ needs to check against type $A$).
One version of that rule would be
\[
  \Infer{changedir-0}
      {
        \Gamma |- e => A
      }
      {\Gamma |- e <= A}
\]
However, the recipe prefers an equivalent rule:
\[
  \Infer{changedir-1}
      {
        \Gamma |- e => A
        \\
        A = B
      }
      {\Gamma |- e <= B}
\]
Since $A = B$ is only equality, rule changedir-1 has exactly the same power
as changedir-0.
Rules changedir-0 and changedir-1 can
be implemented in exactly the same way, but the structure of changedir-1
is closer to that implementation: first, make a recursive call to synthesize $A$;
second, check that $A$ is equal to the given type $B$.
(In logic programming terminology,
in changedir-1 the premise
$\Gamma |- e => A$
has \emph{output freeness}:
the output $A$ is unconstrained, with the constraint imposed later by $A = B$.
In changedir-0, the output $A$ is constrained to be exactly the type from the conclusion.)

The more significant advantage of changedir-1 is that it can be easily extended
to support subtyping.
To turn changedir-1 into a subsumption rule, we only have to replace ``$=$''
with a subtyping relation ``$\subtype$'':
\[
  \Infer{\Sub}
      {
        \Gamma |- e => A
        \\
        A \subtype B
      }
      {\Gamma |- e <= B}
\]
(In a sense, changedir-1 is already a subsumption rule:
equality is reflexive and transitive, so it is a sensible---if extremely limited---subtyping relation.
If we choose equality as the definition of $\subtype$, the rule \Sub is
exactly the rule changedir-1.)

Since $B$ is an input, and $A$ is an output of the first premise,
both $A$ and $B$ are known:
the subtyping judgment can be implemented with both types as input,
with no need to ``guess'' the subtype or supertype.

Subtyping can lead to concerns not addressed solely by the presence of \Sub;
we discuss these concerns in \Sectionref{sec:recipe-subtyping-principal}.

The rule \Sub is not syntax-directed: its subject may be any form of expression.
Because its premise is synthesizing, however, we have some guidance about when
to apply it: when there is a rule whose conclusion can synthesize a type for
that form of expression.
In some bidirectional type systems, such as \citet{Davies00icfpIntersectionEffects},
no expression form has a rule with a synthesizing conclusion and a rule with
a checked conclusion.  In such systems, we can classify the expression forms
themselves as checked or synthesizing, and use \Sub exactly when checking
a synthesizing form.
If we have, say, two rules for \keyword{case} expressions, the situation is more complex.
It is generally best to apply \Sub as late as possible (that is, towards the leaves
of the derivation), because that preserves the information provided by the type
being checked against.

\subsection{Assessing the Recipe}

When we design a bidirectional system according to the recipe,
which criteria are satisfied?

\subsubsection{First criterion: Mode-correctness}
All rules are mode-correct:

\begin{itemize}
\item 
  The variable, annotation and subsumption rules are ``pre-cooked''
  and it is straightforward to verify they are mode-correct.

\item
  In the introduction and elimination rules,
  the judgment directions are chosen to be mode-correct.
\end{itemize}

\subsubsection{Second criterion: Size}
For each type assignment rule,
the recipe produces exactly one rule.
Producing less than one rule is unacceptable, because it would be incomplete
with respect to the type assignment system.

If the original type assignment system did not have a subsumption rule,
the recipe adds one, but this is unavoidable (see the example in \Sectionref{sec:changedir}).
(An alternative would be to duplicate rules and vary the direction of their premises,
but for most type systems, that would lead to more than one additional rule.)

Similarly, the annotation rule is needed to enable a term whose rule has a
checking conclusion to be used in a synthesizing position.
For example, we cannot type the term 
$(\Lam{x} e)e'$
because the first premise of Syn$->$Elim
is synthesizing and the conclusion of Chk$->$Intro is checking.
We need the annotation rule to allow us to type the annotated term
$\big((\Lam{x} e) : A -> B\big)e'$.

Thus, it is not only impossible to remove a single rule, but there is no alternative
approach that can, in general, produce a smaller set of rules.
(We assume that the type assignment system is our ``ground truth'':
we cannot use a prolix type assignment system to argue that
its bidirectionalization is too big.)

\subsubsection{Third criterion: Annotatability}
We write $e' \annoext e$ to mean that $e'$ is a ``more annotated'' version of $e$.
For example, $\annoexp{x}{A} \annoext x$ and $x \annoext x$.
Annotatability says that,
if $\Gamma |- e : A$ (in the type assignment system),
then (1) there exists $e' \annoext e$ such that $\Gamma |- e' <= A$,
and (2) there exists $e'' \annoext e$ such that $\Gamma |- e'' => A$.
We prove this by induction on the type assignment derivation of $\Gamma |- e : A$,
considering cases of the type assignment rule concluding that derivation.

Each type assignment rule has a single corresponding bidirectional rule.
If the conclusion of that bidirectional rule is \emph{checking},
proving part (1) is completely straightforward:
Applying the induction hypothesis to the derivation of each premise
(of the type assignment rule) yields a set of annotated subterms;
combining these annotated subterms gives us our $e'$,
which is typed by the bidirectional rule.
This approach also works if we are
proving part (2) and the conclusion of the bidirectional rule is \emph{synthesizing}.

Going ``into the wind''---proving part (1) with a synthesis rule, or
part (2) with a checking rule---needs only a little more work:

\begin{itemize}
\item 
  If the conclusion of the bidirectional rule corresponding to the type assignment
  rule is \emph{synthesis} and we want to prove part (1), we can show part (2) as above
  to derive
  \[
    \Gamma |- e' => A
  \]
  Now we want to find $e''$ such that $\Gamma |- e'' <= A$.
  Assuming subtyping is reflexive
  (a condition satisfied even by weak subtyping relations, including equality),
  we can derive $A \subtype A$ and use subsumption, giving $\Gamma |- e' <= A$.
  In this case, $e'$ and $e''$ are the same.

\item
  If the conclusion of the bidirectional rule corresponding to the type assignment
  rule is \emph{checking} and we want to prove part (2), we can show part (1):
  \[
    \Gamma |- e'' <= A
  \]
  Now we want to find $e'$ such that $\Gamma |- e' => A$.
  We cannot reuse $e''$, because $\Gamma |- e'' <= A$
  was derived using a checking rule;
  since the recipe produces only one corresponding bidirectional rule,
  we have no rule that can derive $\Gamma |- e'' => A$.
  We must add an annotation:
  \[
     e' ~=~ \annoexp{e''}{A}
  \]
  The last step is to use our annotation rule, deriving $\Gamma |- \annoexp{e'}{A} => A$.
\end{itemize}

\subsubsection{Fourth criterion: Annotation character}
\label{sec:recipe-annotation-character}
Our notion of annotation character (\Sectionref{sec:crit:annotation-character})
posits that annotations should be
(i) lightweight, (ii)--(iii) predictable and stable (with a clear \emph{annotation discipline}), and (iv) legible.

We also argued that a good annotation discipline
should require only local information.
On this point, the recipe does well:
an annotation is required on a subterm
\emph{if and only if}
an introduction form meets an elimination form.

To see why, let's consider how the recipe treats introduction and elimination forms.
Introduction rules type introduction forms, like $\lambda$;
elimination rules type elimination forms, like function application.
Following the recipe, the principal judgment in an elimination form is synthesizing,
so eliminating a variable never requires an annotation.
For example, $f\;x$ needs no annotation, because $f$ synthesizes.
Nor does $(f\;x)\;y$ need an annotation, because $f\;x$ synthesizes.
Variable-binding elimination forms, like \keyword{case},
can also be nested without annotation:
the type of the outer \keyword{case} is propagated to the inner \keyword{case}.
For example, the type $\Int$ is propagated from the conclusion to the inner \keyword{case}:
\[
  \Infer{\SumElim}
     {
       \arrayenvbl{
         \Gamma |- y => (\Bool + \Bool) + \Int
         \\
         \Gamma, x_1 : (\Bool + \Bool)
         |-
         \sumcaseexp{x_1}{x_{11}}{0}{x_{22}}{1} <= \fighi{\Int}
         \\
         \Gamma, x_2 : \Int
         |-
         x_2 <= \Int
       }
     }
     {
       \Gamma |-
       \big(
         \sumcaseexp{y}{x_1}{\sumcaseexp{x_1}{x_{11}}{0}{x_{22}}{1}}{x_2}{x_2}
       \big)
       <=
       \fighi{\Int}
     }
\]
However, we need an annotation at the boundary between introduction and elimination:
in $(\Lam{x} e_1)e_2$,
the introduction form $\lambda$ meets the elimination form $(\cdots)e_2$.
Since the first premise of Syn$->$Elim is synthesizing,
and the conclusion of Chk$->$Intro is checking,
an annotation is needed around $(\Lam{x} e_1)$.

Similarly, in
$\sumcasedots{\injexp{1}{e}}$,
the introduction form \keyword{inj}
meets the elimination \keyword{case},
so $\injexp{1}{e}$ needs an annotation.

In those two examples, we introduced and \emph{immediately} eliminated
the same type ($->$ in the first example and $+$ in the second).
An introduction that is not immediate also requires an annotation.
\[
  \big(
  \sumcaseexp{y}{x_1}{(\Lam{z_1} e_1)}{x_2}{(\Lam{z_2} e_2)}
  \big)
   z
\]
Because the \keyword{case} expression appears as the function part of the application $(\cdots)z$,
it needs to synthesize a type (so we can derive the first premise of Syn$->$Elim).
But the case arms $\Lam{z_1} e_1$ and $\Lam{z_2} e_2$, being introduction forms,
do not synthesize.
Therefore, we need a type annotation around the \keyword{case}, or---more verbosely---%
two annotations, one around each $\lambda$.%
\footnote{In the original recipe, the only option would be an annotation
around the \keyword{case}.
Since the original recipe had only one rule for \keyword{case}, which had a checking conclusion,
annotating the individual arms would have no effect:
typing could not ``pass through'' the \keyword{case} to notice the annotations.}
  Note that if we push the application to $z$ into each arm,
we get a term where eliminations immediately follow introductions
(and, therefore, need annotations):
\[
  \sumcaseexpv{y}
    {x_1}{(\Lam{z_1} e_1)z}
    {x_2}{(\Lam{z_2} e_2)z}
\]

\subsection{Subtyping and Principal Types}
\label{sec:recipe-subtyping-principal}

In this section, we examine subtyping in a bidirectional system with intersection types.
While intersection types are not the most common thing to find in a type system,
they are closely related to parametric polymorphism: 
If a value has type $A_1 \sectty A_2$, it has type $A_1$ and type $A_2$;
if a value has type $\All{\alpha} A$, it has the type $[B/\alpha]A$ for any type $B$.
An intersection type can be seen as a polymorphic type ranging over only two
possibilities; the polymorphic type $\All{\alpha} A$ can be seen as an infinite intersection.

A typing rule is \emph{stationary} \citep[p.\ 55]{Leivant86} if the subject of the premise(s) is
the same as the subject of the conclusion---%
in contrast to (perhaps more familiar)
rules where each premise types a proper subterm of the subject of the conclusion.
In the stationary rules we consider, the subject is \emph{any}
expression $e$. Thus, these rules are not syntax-directed; if the conclusion of a rule types $e$,
the rule is potentially applicable at every step of type checking.

Our rule \Sub is stationary, as are the following rules for intersection types $A \sectty B$
and (implicit) parametric polymorphism $\All{\alpha} A$:
the premises type the same $e$ as the conclusion.
\begin{mathpar}
   \Infer{$\sectty$Intro}
       {
         \Gamma |- e : A_1
         \\
         \Gamma |- e : A_2
       }
       {\Gamma |- e : A_1 \sectty A_2}
   \and
   \Infer{$\sectty$Elim1}
       {
         \Gamma |- e : A_1 \sectty A_2
       }
       {\Gamma |- e : A_1}
   \and
   \Infer{$\sectty$Elim2}
       {
         \Gamma |- e : A_1 \sectty A_2
       }
       {\Gamma |- e : A_2}
   \\
   \Infer{}
       {
         \Gamma, \alpha\,\textsf{type} |- e : A
       }
       {\Gamma |- e : \All{\alpha} A}
   \and
   \Infer{}
       {
         \Gamma |- e : \All{\alpha} A 
         \\
         \Gamma |- B\;\textsf{type}
       }
       {\Gamma |- e : [B/\alpha]A}
\end{mathpar}

With typing rules like these, it makes sense for subtyping to allow
$A_1 \sectty A_2 \subtype A_1$:
the presence of $\sectty$Elim1
means that every term having type $A_1 \sectty A_2$ also has type $A_1$.
Similarly, $A_1 \sectty A_2 \subtype A_2$ because of $\sectty$Elim2.
Observe that both subtyping and Gentzen's rule notation are forms of implication $\imp$:
by treating types as propositions,
$A \subtype B$ becomes $A \imp B$;
a rule is read as
$\textit{premises} \imp \textit{conclusion}$.
So we can translate $\sectty$Elim1:
\[
  ~\arrayenvc{
   \Infer{$\sectty$Elim1}
       {
         \Gamma |- e : \fighi{A_1 \sectty A_2}
       }
       {\Gamma |- e : \fighi{A_1}}
}
   ~~~\text{becomes}~~~
   {A_1 \sectty A_2}
   \imp
   {A_1}
   ~~~\text{becomes}~~~
   \arrayenvc{
     ~\\[-12pt]
   \Infer{}
       { }
       {A_1 \sectty A_2 \subtype A_1}
    }
\]
The rule $\sectty$Elim2 can be treated similarly.

However, the rule $\sectty$Intro cannot be translated in this way:
the two premises mean that the rule cannot be read as ``$\cdots$ implies $\cdots$'',
but only as ``$\cdots$ and $\cdots$ together imply $\cdots$''.
A subtyping judgment $A \subtype B$ can be read as the sequent $A |- B$,
but it is a limited sequent calculus: in addition to allowing only one succedent $B$
(which is a common restriction in sequent calculi),
subtyping allows only one antecedent $A$.
The subtyping rule we would like to construct would need \emph{two} antecedents,
$A_1$ and $A_2$, which don't fit:
\[
  ~\arrayenvc{
   \Infer{$\sectty$Intro}
       {
         \Gamma |- e : \fighi{A_1}
         \\
         \Gamma |- e : \fighi{A_2}
       }
       {\Gamma |- e : \fighi{A_1 \sectty A_2}}
   }
   ~~~\text{becomes}~~~
   {A_1\textsl{\;and\;}A_2}
   ~\imp~
   {A_1 \sectty A_2}
   ~~~\text{becomes}~~~
   \arrayenvc{%
     ~\\[-12pt]
   \Infer{}
       { }
       {A_1\mathsz{14pt}{,\,} A_2 \subtype A_1 \sectty A_2}
    }
\]
The subtyping relation induced
by ``translating''  \emph{only}
the stationary typing rules is weaker (smaller)
than we might desire:
it yields \emph{shallow} subtyping.
For example,
to derive the following through subsumption,
we would need
$(A -> (B_1 \sectty B_2)) \subtype (A -> B_2)$
because
the type of $g$ does not literally match $A -> B_2$.
But the rules for $->$ are not stationary,
letting us forget to add a subtyping rule for $->$:
in the subderivation typing $g$,
we need $g$ to have type $A -> B_2$
but it has only the type $A -> (B_1 \sectty B_2)$.
\[
\inferrule*[Right=$->$Elim]
{
  \dots
  |-
  f : (A -> B_2) -> C
  \\
  \dots
  \not\entails
  g : A -> B_2
}
{
  f : (A -> B_2) -> C,\;
  g : (A -> (B_1 \sectty B_2))
  \not\entails
  f\;g  :  C
}
\]
Note that if we $\eta$-expand $g$ to $\Lam{x} g\;x$,
the term can be typed with only shallow subtyping:
\[
  \inferrule*[Right=$->$Elim]
      {
        \dots |- f : (A -> B_2) -> C
        \hspace{3ex}
        \\
        \inferrule*[Right=$->$Intro]
           {
             \hspace*{-6ex}
             \inferrule*[Right=$\sectty$Elim2]
                {
                  \hspace*{-6ex}
                  \inferrule*[Right=$->$Elim]
                      {
                        \dots, x : A |- g : A -> (B_1 \sectty B_2)
                        \\
                        \dots, x : A |- x : A
                      }
                      {
                        \dots, x : A |- g\,x : B_1 \sectty B_2
                      }
                      \hspace*{-7ex}
                }
                {
                  \dots, x : A |- g\,x : B_2
                }
                \hspace*{-4ex}
           }
           {
             \dots |- (\Lam{x} g\,x) : A -> B_2
           }
           \hspace*{-2.3ex}
      }
      {
        f : (A -> B_2) -> C,\;
        g : (A -> (B_1 \sectty B_2))
        |-
        f\;(\Lam{x} g\,x)  :  C
      }
      \hspace*{14ex}
\]
The technique of $\eta$-expanding to simulate deep subtyping,
\eg for intersection and union types \citep{Dunfield14},
is (as far as we know) due to \citet{Barendregt83};
they showed that putting a $\beta\eta$-expansion rule in
a type system made subsumption admissible (see their Lemma 4.2).

Whether we choose a subtyping relation that is shallow or deep,
we can ``optimize'' a type assignment system
by dropping stationary rules      %
that are encompassed by subsumption.
For example, $\sectty$Elim1 and $\sectty$Elim2 are admissible
by using \Sub with the appropriate subtyping rules.
This engineering optimization is not uniform:
since $\sectty$Intro cannot be translated,
we end up with a type system that has an introduction rule for $\sectty$,
but buries the elimination rules inside subtyping.

Fortunately (if we dislike non-uniform optimizations),
in the bidirectional version of intersection typing,
the bidirectional versions of $\sectty$Elim1 and $\sectty$Elim2 are \emph{not} admissible:
\[
   \Infer{$\sectty$Elim1}
       {
         \Gamma |- e => A_1 \sectty A_2
       }
       {\Gamma |- e => A_1}
   ~~~~~~~
   \Infer{$\sectty$Elim2}
       {
         \Gamma |- e => A_1 \sectty A_2
       }
       {\Gamma |- e => A_2}   
\]
These rules have a synthesizing conclusion,
which means that \Sub cannot simulate them.

It is worth noting, however, that these rules are mode-correct, but not syntax-directed.
With stationary synthesis rules, a term can synthesize many possible types---%
in this example, $e$ can synthesize any of $A_1 \sectty A_2$, $A_1$, and $A_2$.
This is not inherently a bad thing---after all, the whole point of
an intersection type discipline is to allow ascribing many types to
the same term.
However, managing this nondeterminism requires some care both
in the design of the type system, and its implementation. 

\subsubsection{Principal synthesis}

A type assignment system has \emph{principal types} if there always exists a ``best type''---%
a type that represents all possible types.
In systems with subtyping, the best type is the smallest type,
so the principal typing property says that if $e$ is well-typed,
there is a principal type $A$ such that $e$ has type $A$
and $A \subtype B$, for all types $B$ of $e$:

\begin{definition}
\label{def:principal-types}
  A type assignment system has \emph{principal types} if,
  for all terms $e$ well-typed under $\Gamma$,
  there exists a type $A$ such that
  $\Gamma |- e : A$
  and for all $B$ such that $\Gamma |- e : B$,
  we have $A \subtype B$.
\end{definition}

A type inference system has \emph{principal inference} if it implements
a type assignment system that has principal types (\Definitionref{def:principal-types})
and always infers the principal type.  We temporarily write $e => A$ for type inference,
to ease the transition to principal synthesis (\Definitionref{def:principal-synthesis}).
The definition says that if we can infer $A$, then $A$ is principal.

\begin{definition}
\label{def:principal-inference}
  A type inference system has \emph{principal inference} if,
  given $\Gamma |- e => A$,
  for all $B$ such that $\Gamma |- e : B$,
  we have $A \subtype B$.
\end{definition}

That is, every type $B$ that can be assigned to $e$
is a supertype of the inferred type $A$.
(In Damas--Hindley--Milner inference \citep{Hindley69,Damas82},
this property is stated for type schemes;
an inferred type scheme for $e$ is principal when it can be instantiated
to every (monomorphic) type of $e$.)

We can adapt \Definitionref{def:principal-inference} to bidirectional type systems
by using checking, rather than type assignment, to define ``all possible types''.

\begin{definition}
\label{def:principal-synthesis}
 A bidirectional type system has \emph{principal synthesis}
 if, given
  $\Gamma |- e => A$,
  for all $B$ such that $\Gamma |- e <= B$,
  we have $A \subtype B$.
\end{definition}

Principal synthesis is sometimes easy:
if a bidirectional system has uniqueness of both synthesis and checking,
that is, if $\Gamma |- e => A$ and $\Gamma |- e => B$ then $A = B$ (and respectively
for checking),
then $A \subtype B$ because $A = B$.
(In a system in which synthesis always produces the same type $A_1$,
and checking always works against only a single type $A_2$ for a given term,
it had better be the case that $A_1 = A_2$!)

For many sophisticated type systems,
principal synthesis either does not hold or requires some extra design work.
In common formulations of intersection types,
such as the one in \Sectionref{sec:recipe-subtyping-principal},
synthesis and checking are not unique.
For example, if we synthesize $x : (A_1 \sectty A_2) |- x => B$,
the point of the intersection type is to allow either the behaviour $A_1$ or $A_2$,
so it must be possible to derive both
\[
x : (A_1 \sectty A_2) |- x => A_1
\text{~~~~and~~~~}
x : (A_1 \sectty A_2) |- x => A_2
\]
as well as $x : (A_1 \sectty A_2) |- x => A_1 \sectty A_2$.
Saying that synthesis should only produce $A_1 \sectty A_2$ is not
compatible with the recipe:
If $x$ is a function, the rule $->$Elim needs to synthesize a function type;
we are not checking $x$ against a known type, so we cannot rely on subtyping
(which has only a checking conclusion) to eliminate the intersection.

Non-uniqueness means that a straightforward implementation of the rules
must do backtracking search,
trying all three types $A_1$, $A_2$ and $A_1 \sectty A_2$,
even when the choice is irrelevant.
Some backtracking is difficult to avoid with intersection types,
but naively trying all three choices in all circumstances is excessive.

To address this,
the first author's implementation of a bidirectional intersection and union type system
split the synthesis judgment into two:
one judgment that ``maintains principality'',
and one that ``wants an ordinary type''.
The ``maintains principality'' judgment lacked rules like $\sectty$Elim1 and $\sectty$Elim2;
the ``ordinary type'' judgment included such rules.
The choice of synthesis judgment depended on the rule.
The premise of a \keyword{let} rule, synthesizing a type for the let-bound expression,
used the ``maintains principality'' judgment to ensure that
the variable typing added in the body of the \keyword{let} was principal.
So, for example, if the let-bound expression was simply a variable $x$
of type $A_1 \sectty A_2$, the typing in the body would also have $A_1 \sectty A_2$,
with no backtracking between choices of $\sectty$Elim1 and $\sectty$Elim2.
However, the premise of $->$Elim used the ``wants an ordinary type'' judgment,
because we may need to apply rules like $\sectty$Elim1 to expose the $->$ connective.
See \citet[\S 6.7.1 on pp.\ 186--187]{DunfieldThesis}.

\citet[\S 2.10.2]{DaviesThesis} includes a bidirectional typing (\emph{sort checking})
system with a principal synthesis property.
It appears that Davies asserts the property (page 41) without formally stating or proving it,
but from our reading of the rules on page 42, it holds as follows:
Principal synthesis is achieved through an auxiliary judgment that,
when applying a function of type
$(R_1 -> S_1)
\sectty
\dots
\sectty
(R_n -> S_n)$,
gathers all the components $R_i -> S_i$
such that the function argument checks against $R_i$,
and synthesizes the intersection of all such $S_i$.
(\citet[\S 3.10]{DaviesThesis} also discusses a principal sorts property
in a non-bidirectional type inference setting,
but this is less relevant to our survey.)

\section{Polymorphism}
\label{sec:polymorphism}

Damas--Milner type inference \citep{Damas82}
allows only prefix polymorphism:
the quantifiers in a type must appear on the outside,
allowing $\All{\alpha} (\All{\beta} \alpha -> \beta -> \alpha)$
but not $\All{\alpha} \alpha -> (\All{\beta} \beta -> \alpha)$
and $\All{\beta} (\All{\alpha} \alpha -> \alpha) -> \beta -> \beta$.
This restriction is called \emph{prefix} or \emph{prenex} polymorphism.
In their terminology, types contain no quantifiers at all;
only type \emph{schemes} can have quantifiers (on the outside).
Polymorphism can be introduced only on \textkw{let} expressions.

If programs must indicate
both where and how to introduce and eliminate polymorphism,
polymorphism is \emph{fully explicit}.
Adding fully explicit polymorphism to a bidirectional system is straightforward:
since the term is an input,
both the introduction and elimination rules can use the information in the term.
However, fully explicit polymorphism is often considered unusable,
mostly because of the explicit eliminations:
it is burdensome to say how to instantiate every quantifier.

Explicit \emph{introduction} of polymorphism can readily cope with
less restrictive forms of polymorphism: \emph{higher-rank} polymorphism,
which allows quantifiers to be nested anywhere in a type (including to the left of arrows,
as in the type
$\All{\beta} (\All{\alpha} \alpha -> \alpha) -> \beta -> \beta$
mentioned above),
and even \emph{impredicative} polymorphism which allows quantifiers to be instantiated
with polymorphic types.

Making eliminations implicit is not easy.
Following the recipe, eliminations synthesize,
so given $e => (\All{\alpha} \alpha -> \alpha)$
we should derive $e => [\tau/\alpha](\alpha -> \alpha)$,
that is, $e => (\tau -> \tau)$.
Unfortunately, the instantiation $\tau$ (we write $\tau$, rather than $A$, for \emph{monotypes}---%
types containing no quantifiers) is decidedly not an input.

This \emph{instantiation problem} has been tackled from several directions.
The first widely known paper on bidirectional typing
\citep{Pierce00} considered the problem in a setting with subtyping,
and answered it by local constraint solving
(hence the title \emph{Local Type Inference})
around function applications.
Subtyping leads to considering the upper and lower bounds of a type,
rather than equations on types.
Pierce and Turner restricted instantiation to prefix polymorphism,
though their source language allowed impredicative polymorphism
if the programmer explicitly instantiates the quantifier.

Their bidirectional rules are quite different from what our
recipe might produce in their setting.
They have three rules for $\lambda$-abstractions:
S-Abs synthesizes the type of a $\lambda$ with an explicit argument type,
C-Abs-Inf checks a $\lambda$ without an explicit argument type,
and C-Abs checks a $\lambda$ with an explicit argument type.
They have two rules for applications with explicit quantifier instantiation:
S-App synthesizes, with the same directionality as our $->$-elimination rule,
and C-App checks the application against a type $U$
by synthesizing a type $S -> R$ for the function
and checking that $R$ is a subtype of $U$.
(Here, we elide substituting the explicit instantiation type $T$.)
Applications with inferred instances also have two rules:
S-App-InfSpec synthesizes,
with a premise that forces the chosen instantiation to
produce the best synthesized result type;
C-App-UInfSpec checks, and lacks that premise because
the needed result type of the application is known.
\citet{Hosoya99:LocalTypeInf} discuss the annotation burden of
local type inference for several example programs.

One can argue that \emph{all} type systems have subtyping,
where some systems have only trivial subtyping
($A$ is a subtype of $B$ iff $A = B$).
A more moderate perspective is that ``most'' type systems have subtyping:
even in prefix polymorphism,
types that are ``more polymorphic'' can be considered subtypes.
By the substitution principle of \citet{Liskov94},
$\All{\alpha} \alpha -> \alpha$
should be a subtype of
$\unitty -> \unitty$:
any program context that expects an identity function on $\unitty$---%
of type $\unitty -> \unitty$---%
should be satisfied by a polymorphic identity function
of type $\All{\alpha} \alpha -> \alpha$.
(In many systems, including Damas--Milner, types cannot contain
quantifiers---only type schemes can---%
but the perspective could be adapted to subtyping on type schemes,
and is conceptually useful in any case.)
In systems with higher-rank polymorphism,
the perspective that polymorphism is a form of subtyping is salient:
since quantifiers can appear to the left of arrows,
we may want to pass a ``more polymorphic'' argument to a function
that expects something less polymorphic.

\subsection{``Complete and Easy'' Polymorphism}
\label{sec:complete-and-easy}

In this subsection, we explain the key elements of
our technique \citep{Dunfield13}, discuss some typing rules,
and describe its history in more detail.

\subsubsection{Greedy instantiation}

The key idea taken from \citet{Cardelli93:FsubTheSystem} was that,
when eliminating a polymorphic type,
we can treat the first plausible solution as \emph{the} solution.
For example, if we are calling a function of type
$\All{\alpha} \alpha -> \alpha -> \alpha$
(assuming parametricity, such a function can only return one of its arguments)
and pass as the first argument something of type \tyname{Cat},
we instantiate $\alpha$ to \tyname{Cat}.
This works perfectly well when the second argument has the same type,
or when the second argument is of a subtype of \tyname{Cat}
(\eg \tyname{Tabby}),
but fails when the second argument is of a larger type.
If the first argument has type \tyname{Tabby}
but the second argument has type \tyname{Cat},
the second argument will fail to check against \tyname{Tabby},
since not all cats are tabbies.

In its original setting, this ``greedy'' method's
vulnerability to argument order was rather unfortunate.
In a setting of predicative higher-rank polymorphism
without other forms of subtyping, however, it can work nicely.
The ``tabby-first problem'' cannot arise because the only way a type
can become strictly smaller is by being strictly more polymorphic,
and if the first argument is polymorphic we would be instantiating $\alpha$
with a polymorphic type, which would violate predicativity.

\subsubsection{Systems and judgments}

The paper focused on two bidirectional type systems:
a \emph{declarative} system
whose $\forall$-elimination rule ``guesses'' types,
and an \emph{algorithmic} system which instead uses greedy instantiation.

Our declarative system followed a looser version of the Pfenning recipe:
in addition to the rules produced by the recipe,
the declarative system included synthesizing introduction rules for $\unitty$ and $->$.
A subsumption rule, \rulename{DeclSub}, used a declarative subtyping relation $\leq$
whose ``$\forall$-left'' rule---the rule concluding $(\All{\alpha} A) \leq B$---%
guessed a monotype $\tau$ to use in the premise $[\tau/\alpha]A \leq B$.

We also incorporated an \emph{application judgment},
written
\[
  \declappjudg{\Psi}{A}{e}{C}
\]
meaning that under the declarative context $\Psi$,
if a function \emph{of type} $A$ is applied to an argument $e$,
the entire application will have result type $C$.
The rules for this judgment eliminate $\All{\alpha} A$ by
substituting an unsolved type variable $\ahat$ for $\alpha$.

\subsubsection{Ordered typing contexts}

Rather than passing along a ``bag of constraints'',
we can store the (solved and unsolved) type variables (written $\ahat$, $\bhat$, etc.)
in an \emph{ordered} context.
Issues of circularity and scope still need care,
but the way to handling them is clarified:
if $\ahat$ appears to the left of $\bhat$
and we need to constrain them to be equal,
we must solve $\bhat$ to $\ahat$, not the other way around.
Forcing this single order avoids backtracking.

In our algorithmic system, the three typing judgments---checking, synthesis and application---%
included an \emph{output context} $\Delta$.
For example, if the \emph{input} context $\Gamma = (\ahat, x : \ahat)$,
meaning that $\ahat$ is an unsolved existential variable and $x$ has type $\ahat$,
checking $x$ against $\unitty$ will solve $\ahat$:
\[
  \ahat, x : \ahat |- x <= \unitty -| \hypeq{\ahat}{\unitty}, x : \ahat
\]
More generally, in a derivation of $\Gamma |- \cdots -| \Delta$,
the output context $\Delta$ \emph{gains information}:
any solutions present in $\Gamma$ are also present in $\Delta$,
but unsolved $\ahat$ in $\Gamma$ may gain solutions in $\Delta$.

We made this idea of information gain precise by defining \emph{context extension}:
whenever a judgment $\Gamma |- \cdots -| \Delta$ is derivable,
the output context $\Delta$ is an extension of $\Gamma$,
written $\substextend{\Gamma}{\Delta}$.
As in the $x <= \unitty$ example, information about existential type variables may increase in $\Delta$;
also, new existential variables (unsolved or solved) may appear in $\Delta$.
However, the ``ordinary'' program variable typings $x : A$ must not change.

\subsubsection{Contexts as substitutions}

In our paper, we allowed contexts to be used as substitutions:
if $\Delta$ contains $\hypeq{\ahat}{\unitty}$, then $[\Delta]\ahat = \unitty$.
This usage pervaded the system.
For instance,
the subsumption rule applies $\Theta$, the output context of the first premise,
to the inputs in the second premise:
\[
     \Infer{\Sub}
          {\synjudg{\Gamma}{e}{A}{\Theta}
            \\
            \subjudg{\Theta}{[\Theta]A}{[\Theta]B}{\Delta}
          }
          {\chkjudg{\Gamma}{e}{B}{\Delta}}
\]
Such applications---%
found in all our rules with more than one premise---%
guarantee that
whenever the input types in a judgment do not contain existential variables
already solved in the input context,
the output types do not contain existential variables that are solved in the \emph{output} context.
That is, all types are ``solved'' as much as possible.
While this property made the rules a little more complicated,
it seemed to make the system easier to work with.

\subsubsection{Historical notes and other approaches}

The first author combined the two key elements,
greedy instantiation and ordered contexts,
in a workshop paper \citep{Dunfield09};
the idea of using ordered contexts is due to Brigitte Pientka.
Unfortunately, key proofs in the paper were severely flawed.
Despite these flaws, the second author liked the key ideas
and wanted to use them in a type system with higher kinds
and inverse types.
We haven't built that system yet, but the ``preliminary'' step of shoring up
the workshop paper became its own paper \citep{Dunfield13}.
\citet{Gundry10} offer a reformulation of Algorithm W
that is based on information increase over ordered contexts.

We know of several languages that have used our approach or a variant of it:
Discus%
\footnote{\url{http://blog.discus-lang.org/2017/10/the-disciplined-disciple-compiler-v051.html}},
PureScript%
\footnote{\url{http://www.purescript.org/}},
and Hackett%
\footnote{\url{https://github.com/lexi-lambda/hackett}}.
\citet{Xie18:ConsistentSubtypingForAll} extended the approach and its metatheory;
their work is discussed in \Sectionref{sec:gradual-typing}.

The first implementation of higher-rank polymorphism to see widespread
use was in GHC Haskell, and was documented in
\citet{PeytonJones07}. The techniques introduced in this paper were
quite similar to ours~\citep{Dunfield13}, but as usual we did not
understand the closeness of the relationship until after we
reinvented our own version.
They specified their algorithm in the Hindley--Milner style:
in their specification, variables are automatically instantiated when used and re-generalized as needed, via generalization and specialization rules that are not syntax-directed.
Their algorithm eagerly instantiates quantifiers, re-generalizing let-bound expressions; it uses subtyping (specialization) only at bidirectional mode switches.
Our algorithm is lazy, never instantiating a quantifier unless it has to.
\citet{Eisenberg16:visible-type-app} extended the algorithm of \citet{PeytonJones07}
with explicit quantifier instantiations.
Implementing this approach required lazy instantiation so that the algorithm could delay instantiation until user applications had been processed.

\citet{Zhao19} give a new machine-checked formalization of type
inference, using our declarative specification~\citep{Dunfield13}.
However, they give a new algorithm for easier machine verification,
in a ``worklist'' style:
Type inference is broken into subproblems %
just as with other bidirectional systems, but instead of writing the
type inference algorithm as a simple recursion on the structure of the
syntax, the problems are pushed onto a stack as the syntax is decomposed.
This technique avoids the need for an output context---as type checking
refines the values of the existential variables, this is automatically
propagated to all the remaining subproblems.
By associating contexts with judgments on the worklist,
rather than threading a context through the derivation,
they avoid the need for our ``scope markers''.
The larger tradeoffs are not entirely clear, but we can say that
our algorithm solves problems in a more predictable order,
while
their approach is potentially more powerful:
it enables the algorithm to defer solving problems until more information is available.

\subsection{Extensions to Polymorphism}
\label{sec:extensions-to-polymorphism}

\citet{PeytonJones07}, our 2013 system, and \citet{Zhao19}
support predicative polymorphism, where quantifiers can be instantiated
only with monotypes.  
Inferring instantiations for impredicative polymorphism
is undecidable in general.
Building on the approach of \citet{PeytonJones07}, \citet{Serrano20}
support \emph{guarded} impredicativity, where the quantifier occurs
inside a type constructor.

Another extension to polymorphism is generalized algebraic datatypes (GADTs),
in which datatypes are not uniformly polymorphic:
the type arguments can depend on the constructor \citep{Xi03:guarded}.
\citet{Pottier06:stratified} use bidirectional typing to produce GADT-related annotations,
which provide the missing information for non-bidirectional type inference.
OutsideIn \citep{Vytiniotis11} and \citet{Dunfield19} also use bidirectional
typing for GADTs.  Both use unification to propagate equality information.
Our paper has a declarative specification, though our proofs that our algorithm
is sound and complete are quite involved.
OutsideIn is more liberal about when it can
unify two types without an annotation, which makes it more powerful than our system
However, formulating a specification for OutsideIn remains an open problem.

\section{Variations on Bidirectional Typing}
\label{sec:variations}

When discussing work that uses different notation from ours, \eg $\downarrow$ and $\uparrow$
instead of $<=$ and $=>$, we replace the original notation with ours.
See \Sectionref{sec:notation}.

\subsection{Mixed-direction Types}

Instead of distinguishing checking from synthesis at the judgment level,
\citet{Odersky01:ColoredLocal}
make a distinction in type syntax:
(1) \emph{inherited} types
$\oderskyinh{A}$
serve the purpose of the checking judgment,
and
(2) \emph{synthesized} types
$\oderskysyn{A}$
serve the purpose of the synthesis judgment.
In their system,
\emph{general types} combine inherited and synthesized types.
For example,
in $\oderskyinh{(\oderskysyn{\Int} -> \oderskyinh{\Bool})}$
the outermost $\oderskyinhsym$ denotes that the connective $->$ is inherited (in our terminology, checked),
the $\oderskysynsym$ that precedes $\Int$
denotes that the domain of the function is synthesized,
and the $\oderskyinhsym$ that precedes $\Bool$
denotes that the range of the function is inherited (checked).
Their subtyping judgment does not synthesize nontrivial supertypes:
$\oderskyinh{\Int} \subtype \oderskysyn{\topty}$
is \emph{not} derivable,
but
$\oderskyinh{\Int} \subtype \oderskysyn{\Int}$
is derivable
(the supertype is trivial, being equal to the subtype).
When the supertype is inherited (checked), as in $\oderskyinh{\Int} \subtype \oderskyinh{\topty}$,
subtyping for Odersky \etal corresponds to the bidirectional subsumption rule.

Another approach to pushing these distinctions into the type syntax
can be found in the work on \emph{boxy types}~\citep{boxy-types}. The
type syntax of boxy types is simpler than in the work of
\citet{Odersky01:ColoredLocal}, since it does not permit arbitrary
interleaving of checked and synthesized type components of a type:
inferred types occur in boxes, and boxes are not allowed to nest.
In addition, the treatment of variables does not follow the
basic bidirectional recipe; instead, variables are \emph{checked},
which is more similar to the backwards approach to bidirectional
typing we discuss in \Sectionref{sec:backwards}. 

\subsection{Directional Logic Programming}

One lesson that can already be drawn is that the flow of information
through the typing judgments is a key choice in the design of
bidirectional systems, and that it is often desirable to go beyond the
simple view of modes as either inputs or outputs: for example,
\citet{Odersky01:ColoredLocal} track whether each part of a type is an
input or output. So describing bidirectional typing algorithms
can require a more subtle notion of mode. 

\citet{Reddy93} adapts ideas from classical linear logic to
characterize \emph{directional} logic programs.  Directional logic
programming subsumes moded logic programming. For example, a ternary
predicate $p$ in regular multi-sorted predicate logic might be given
the type:
\[
  p : \texttt{List}(\texttt{Int}) \times \texttt{Int} \times \texttt{Bool} \to \mathsf{prop}
\]
This says that a proposition of the form $p(X, Y, Z)$ has three
arguments, with $X$ a list of integers, $Y$ an integer, and $Z$ a
boolean. With an ordinary mode system, each of these three arguments
must be classified entirely as an input or output.

However, in directional logic programming, modes become part of the
structure of the sorts, which permits giving sorts like:
\[
  p : \texttt{List}(\texttt{Int}^\bot) \tensor \texttt{Int} \tensor \texttt{Bool}^\bot \to \mathsf{prop}
\]
Now, in a predicate occurrence of the form $p(X, Y, Z)$, the
argument $Y$ is an input integer, and $Z$ is an output boolean, with the
an output of sort $\tau$ marked as $\tau^\bot$. The  argument $X$
has the sort 
$\texttt{List}(\texttt{Int}^{\botty})$, meaning that the list is
structurally an input (so its length is known) but its elements are
outputs: the type $\texttt{Int}$ denotes an integer that is an input,
while $\texttt{Int}^{\botty}$ denotes an integer that is an output. 

The notation $A^{\botty}$ corresponds to negation, and the classical
nature of the sort structure arises from the fact that outputting
an output $A^{\bot\bot}$ is the same as an input $A$---i.e., the user
must supply a box which will be filled with an $A$.

This more fine-grained structure lets us give sort declarations which capture
the fact that (for example) the boxes in boxy types are outputs but the rest of
the type is an input.

\subsection{Mode Annotations}
\citet[pp.\ 242--243]{DaviesThesis} describes \emph{mode annotations}
that would allow programmers to declare
which functions should be typed using synthesis (instead of checking)
and whether the entire application should be checking (instead of synthesizing).
As far as we know, mode annotations were never implemented.

Davies motivated this annotation form for polymorphic functions like ``higher-order case'',
which takes as arguments an instance of a datatype (\texttt{bits})
and a series of functions corresponding to case arms.
That is, the ``first-class'' case expression
\[
  \caseexp{e_{\text{bits}}}
          {e_{\text{bnil-case}}
            \matchor
            x.\,e_{\text{b0-case}}
            \matchor
            y.\,e_{\text{b1-case}}
          }
\]
is written
$
\texttt{bcase}\;%
{e_{\text{bits}}}\;%
(\lam{\unitexp} e_{\text{bnil-case}})\;%
(\lam{x} e_{\text{b0-case}})\;%
(\lam{y} e_{\text{b1-case}})%
$.

We can declare the type and \emph{mode} of the function \texttt{bcase}:
\[
\tabularenvl{
\textkw{val}
\texttt{
bcase
:
bits $->$ (unit $->$ $\alpha$) $->$ (bits $->$ $\alpha$) $->$ (bits $->$ $\alpha$) $->$ $\alpha$
}
\\
\textkw{mode}%
\texttt{
bcase
:
}\textit{inf} $->$ \textit{chk} $->$ \textit{chk} $->$ \textit{chk} $->$ \textit{chk}
}
\]
The second line is a mode annotation.
It says that the first argument of \texttt{bcase}
should synthesize (be \textit{inf}\hspace*{-0.13ex minus 0.15ex}erred),
the remaining three arguments should be checked,
and the entire application should be checked (the last \textit{chk}).
Synthesizing the type of the first argument corresponds
to the synthesizing principal judgment of the elimination rule \SumElim;
checking the other arguments corresponds to the checking premises of \SumElim;
checking the entire application corresponds to the checking conclusion of \SumElim.

One can view mode annotations as instructions for transforming a type $A -> B$
into a ``decorated'' type---%
something like \citet{Odersky01:ColoredLocal},
but where the connective itself is decorated.
The recipe's elimination rule $->$Elim
would correspond to a type
${}^{=>}{->}^{<=}_{=>}$,
matching the scheme 
${}^{\textit{pr1}}{->}^{\textit{pr2}}_{\textit{conc}}$
where \textit{pr1} is the first premise,
\textit{pr2} is the second premise
and \textit{conc} is the conclusion.
Not all such decorations would be mode correct.

\subsection{Simultaneous Input and Output}
\label{sec:simultaneous}

Another way to blend input and output was developed
by \citet{Pottier06:stratified} for type inference for GADTs \citep{Xi03:guarded}.
The overall structure of their system is unusual (their bidirectional algorithm
produces \emph{shapes}, which are then given to a non-bidirectional type inference
algorithm); for brevity, we explain the idea in a more ordinary setting
involving types, not shapes.
The basic idea is to
combine synthesis and checking judgments into
a single judgment with two types:
$\Gamma |- e <= A => B$
means ``check $e$ against $A$, synthesizing $B$'',
where $B$ is a subtype of $A$.
The system thus operates in both checking and synthesis modes simultaneously.

A similar idea is used in the program synthesis work of \citet{Polikarpova16},
\emph{round-trip type checking}:
combines a checking judgment
$\Gamma |- e <= A$
with a \emph{type strengthening} judgment
$\Gamma |- e <= A => B$.

For example, the strengthening rule for variables 
\citep[Fig.\ 4]{Polikarpova16}
checks against a given type $A$,
but utilizes $\Gamma$'s refinement type
$\{b \such \psi\}$
(base type $b$ such that the constraint $\psi$ holds)
to produce a strengthened type
$\{b \such \nu = x\}$.
\[
  \Infer{\textsc{VarSc}}
      {
        \Gamma(x) = \{b \such \psi\}
        \\
        \Gamma |- \{b \such \psi\} \subtype A
      }
      {
        \Gamma |- x <= A => \{b \such \nu = x\}
      }
\]
Synthesis becomes a special case of strengthening:
$\Gamma |- e => B$
can be written
$\Gamma |- e <= \textsf{top} => B$.
Since every type is a subtype of \textsf{top},
the synthesized type $B$ is stronger than the goal type (\textsf{top}).

\subsection{Backwards Bidirectional Typing}
\label{sec:backwards}

In the basic Pfenning recipe,
the principal judgment in an introduction rule is checked,
and the principal judgment in an elimination rule synthesizes.
However, \citet{Zeilberger15} observed
that in a multiplicative linear type theory,
bidirectional typing works precisely as well if you did it backwards,
changing all occurrences of synthesis to checking, and vice versa.
Zeilberger's observation was made in the context of a theorem
relating graph theory and type theory,
but it is a sufficiently striking result that it is worth spelling out in its own right.
We will not precisely replicate his system,
but we will discuss our divergences
when relating it to other bidirectional type systems.

First, we give the syntax of multiplicative linear logic.
\begin{grammar}
{Types} & $A$ & \bnfas & $1 \bnfalt A \tensor B \bnfalt A \lolli B$
\\
{Terms} & $e$ & \bnfas & $x \bnfalt \lam{x}{e} \bnfalt e\,e'
\bnfaltBRK
\unit \bnfalt \letunit{e}{e'}
\bnfaltBRK \pair{e}{e'} \bnfalt \letpair{x}{y}{e}{e'}$
\\
{Contexts} & $\Gamma$ & \bnfas & $\cdot \bnfalt \Gamma, x \From A$
\end{grammar}
The types of MLL are
the unit type $1$,
the tensor product $A \tensor B$,
and the linear function space $A \lolli B$.
Unit and tensor are introduced by $\unit$ and $\pair{e}{e'}$,
and are eliminated by pattern matching.
Functions are introduced by $\lam{x}{e}$
and eliminated using applications $e\,e'$.

Contexts are a bit unusual---they pair together variables and their
types as usual, but instead of treating a variable as a placeholder
for a \emph{synthesizing} term,
we treat variables as placeholders for \emph{checking} terms.
This will have substantial implications for the
mode discipline of the algorithm, but we will defer discussion of this
point until the whole system is presented. 

Now we give the typing rules, starting with those for the unit type.
\begin{mathpar}
\Rule{ }
    { \synth{\cdot}{\unit}{1} }
\and 
\Rule
  {
    \synth{\Delta}{e'}{A}
    \\
    \CHECK{\Gamma}{e}{1} 
  }
  {
    \synth{\Gamma, \Delta}{\letunit{e}{e'}}{A}
  }
\end{mathpar}
The introduction rule says that in an empty context, the unit value
$\unit$ \emph{synthesizes} the type $1$.
The pattern-matching style elimination $\letunit{e}{e'}$
first synthesizes a type $A$ for the body $e'$,
and then checks that the scrutinee $e$ has the unit type $1$.

Thus, we synthesize a type for the continuation first, before checking the
type of the data we are eliminating;
this is the exact reverse of the Pfenning recipe.
For the unit type, this is a mere curiosity, but it gets more interesting with the tensor product
type $A_1 \tensor A_2$.
\begin{mathpar}  
\Rule{ \synth{\Gamma}{e_1}{A_1} \\ 
       \synth{\Delta}{e_2}{A_2} }
     { \synth{\Gamma, \Delta}{\pair{e_1}{e_2}}{A_1 \tensor A_2} }
\and
\Rule
  {
    \synth{\Gamma, x_1 \From A_1, x_2 \From A_2}{e'}{C}
    \\
    \CHECK{\Delta}{e}{A_1 \tensor A_2}
  }
  {
    \synth{\Gamma, \Delta}{\letpair{x_1}{x_2}{e}{e'}}{C}
  }
\end{mathpar}
The synthesis rule for pairs remains intuitive,
though it reverses the direction given by the Pfenning recipe:
for a pair $\pair{e_1}{e_2}$,
we first synthesize $A_1$ for $e_1$
and $A_2$ for $e_2$,
then conclude that the pair
has type $A_1 \tensor A_2$.

However, the elimination rule typing $\letpair{x_1}{x_2}{e}{e'}$ is startling.
First, it synthesizes the type $C$ for the continuation
$e'$; we learn from having typed $e'$
that $x_1$ and $x_2$ need to have types $A_1$ and $A_2$ respectively.
This gives us the information we need to check $e$ against $A_1 \tensor A_2$.
The linear function type $A_1 \lolli A_2$ has a similar character:
\begin{mathpar}
\Rule{ \synth{\Gamma, x \From A}{e}{B} }
     { \synth{\Gamma}{\lam{x}{e}}{A \lolli B} }
\and
\Rule{ \synth{\Gamma}{e'}{A} \\ 
       \CHECK{\Delta}{e}{A \lolli B} }
     { \CHECK{\Gamma, \Delta}{e\,e'}{B} }
\end{mathpar}
Here, to synthesize a type for the introduction form $\lam{x}{e}$, we
synthesize $B$ for the body $e$, and then look up what type
$A$ the argument $x$ needs to have in order for the body $e$ to be well typed.
To check that an application $e\,e'$ has the type $B$,
we synthesize $A$ for the argument $e'$,
and then check the function $e$ against $A \lolli B$.

The rule for product elimination suggests a reversed rule for let-expressions,
which allows us to defer checking the bound expression $e$ until we know
what type it needs to have.
\begin{mathpar}
   \Infer{}
      {
        \CHECK{\Gamma, x \From A}{e'}{B}
        \\
        \CHECK{\Gamma}{e}{A}
      }
      {
        \CHECK{\Gamma}{\letv{x}{e}{e'}}{B}
      }
\end{mathpar}

Again, the checking/synthesis modes are reversed from
most bidirectional type systems.
We can see how this reversal plays out for variables below:
\begin{mathpar}
\Rule{  }
     { \CHECK{x \From A}{x}{A} }
\and
\Rule{ \synth{\Gamma}{e}{A} \\ A = B}
     { \CHECK{\Gamma}{e}{B} }
\end{mathpar}
Here, when we check that the variable $x$ has type $A$, the
context must be such that it requires $x$ to have the type $A$.
However, the switch between checking and synthesis is standard.

Relative to most bidirectional systems, the information flow in the
variable rule (as well as for pattern matching for pairs and
lambda-abstraction for functions) is strange.
Usually, the context would give the type of each variable.
However, in this case the context \emph{is told} the type of each variable.
This system of rules is still well-moded in the logic programming sense,
but the moding is more exotic than simple inputs or outputs.
Within a given context, the variables are inputs, but their types are outputs.
Following \citet{Reddy93}, the moding of checking and synthesis might be given as
\[    %
\begin{tabular}{l}
\texttt{ mode check : (List (Var $\tensor$ Type$^{\botty}$) $\tensor$ Term $\tensor$ Type)} $\to \mathsf{prop}$
\\
\texttt{ mode synth : (List (Var $\tensor$ Type$^{\botty}$) $\tensor$ Term $\tensor$ Type$^{\botty}$)} $\to \mathsf{prop}$
\end{tabular}
\]
This mode declaration says that for both checking and synthesis,
the spine of the context and the variable names are inputs,
but the ascribed type for each variable is an output.
Similarly, the term is an input in both judgments,
but the type is an input in \texttt{check} but an output in \texttt{synth}.

We can relatively easily prove a substitution theorem for the backwards system:
\begin{theorem}{(Backwards Substitution)}
\label{thm:backwards-substitution}
If $\CHECK{\Delta}{e}{A}$, then
\begin{enumerate}
\item If $\CHECK{\Gamma, x \From A, \Theta}{e'}{C}$ then $\CHECK{\Gamma, \Delta, \Theta}{[e/x]e'}{C}$.
\item If $\synth{\Gamma, x \From A, \Theta}{e'}{C}$ then $\synth{\Gamma, \Delta, \Theta}{[e/x]e'}{C}$
\end{enumerate}
\end{theorem}

Unfortunately, we do not presently know how to nicely characterize
the set of terms that is typable under this discipline,
unlike the characterization for the Pfenning recipe that
annotation-free terms are the $\beta$-normal terms.

\subsubsection{Applications of Backwards Bidirectional Typing}
\label{sec:arguments-first}

When designing a bidirectional type system, what approach should we use---the
Pfenning recipe, its pure reversal (as presented above), or something else?
The most popular answer seems to be ``something else'':
Many practical systems synthesize types for literals (unit, booleans, etc.)
and for pairs, which---being introduction forms---can only be checked
under the strict Pfenning recipe.
However, a number of papers have used reversed rules in more subtle ways.
Which approach to take depends on the choice of tradeoffs:
for example, we can require fewer annotations if we complicate the flow of
type information.

Drawing inspiration from relevance logic, which requires variables to be used
at least once (as opposed to the exactly-once constraint of linear logic),
\citet{Chlipala05} require a let-bound variable to be used at least once.
This allows them to reverse the typing rule for let-expressions,
reducing the annotation burden of the basic Pfenning recipe:
no annotation is needed on the let-bound expression.
Their typing contexts contain checking variables, whose moding
is similar to the variables in our backwards bidirectional system.
Such a variable must occur at least once in a checking position;
that occurrence determines the type of the variable,
which can be treated as synthesizing in all other occurrences.

\citet{Xie18:LetArgumentsGoFirst} present another bidirectional type
system for polymorphism. Their rule for function application is very similar to
the backwards rule presented here, with the idea that backwards
typing means that applications like $(\fun{x}{e})\;e'$ do not
need a type annotation at the redex. This requires fewer annotations
on let-bindings, as in the work of \citet{Chlipala05},
but with support for polymorphism.

\citet{Zeilberger15} (and its follow-up work \citet{Zeilberger18}) did
not use any type annotations at all.  Instead, his bidirectional
system was used to deduce a type scheme for the linear lambda terms,
in the style of ML type inference, to find a simple proof of the fact
every linearly typed term has a most general type, and moreover that
the structure of its $\beta$-normal, $\eta$-long form
is determined by this type scheme.

Intersection types can reconcile multiple occurrences of the
same variable at different types; it appears that
the type inference algorithm of \citet{DolanThesis} can be viewed as
calculating intersections via a computable lattice operation on types.

\section{Proof Theory, Normal Forms, and Type Annotations}
\label{sec:proof-theory}

\subsection{Subformula Property}
\label{sec:subformula-property}

In cut-free sequent calculi,
every formula (proposition) that appears in a derivation
is a subformula of some formula in the conclusion.
For example, in the following sequent calculus derivation,
the formulas
$(P \land Q) \land R$
and
$P \land Q$
are subformulas (subterms)
of the conclusion's assumption
$(P \land Q) \land R$.
\[
  \inferrule
      {
        \inferrule
           {
             \inferrule
                 { }
                 {(P \land Q) \land R |- (P \land Q) \land R}
           }
           {
             (P \land Q) \land R |- P \land Q
           }
      }
      {
        (P \land Q) \land R |- Q
      }
\]
Through the Curry--Howard correspondence,
a property of formulas becomes a property of types,
but the property is still called the \emph{subformula} property, to avoid confusion with ``subtype''.%

A consequence of the subformula property is that
if a connective appears in a formula, such as the $\land$ in $P \land Q$ in the middle step,
the connective must appear in the conclusion.
This consequence is useful in a number of type systems, because it ensures that
problematic type connectives appear only with the programmer's permission.
Bidirectional type systems based on the recipe in \Sectionref{sec:recipe}
only synthesize types that are (subformulas of) annotations:
to eliminate an $->$, the function subterm must synthesize by reason of
being a variable, an annotation, or an elimination form.
The type of a variable flows from an annotation on the binding form
(\eg a $\lambda$ inside an annotation)
or from the synthesizing premise of an elimination rule
(\eg the premise typing the scrutinee of a \keyword{case}).

Consider, for example, intersection and union types.
Efficient type-checking for intersections and unions is difficult
\citep{Reynolds96:Forsythe,DunfieldThesis};
intersections and unions that come ``out of nowhere'' in the middle of a derivation---%
without being requested via a type annotation---%
would aggravate this difficulty.
Another example is found in type systems that encode multiple evaluation strategies:
if a programmer generally prefers call-by-value, but occasionally wants to use call-by-name,
the subformula property implies that call-by-name connectives
appear only when requested \citep{Dunfield15}.
Risky connectives abound in gradual type systems:
\emph{unknown} or \emph{uncertain} types
should appear only with the programmer's permission,
because they permit more dynamic failures than other type connectives do \citep{Jafery17}.

In type inference, all of typing is in a single judgment.
Without a checking judgment, there is no goal type;
to increase typing power,
one must put more and more ``cleverness'' into inference.
Certain kinds of cleverness destroy the subformula property:
automatic generalization, for example, creates ``for all'' connectives out of nowhere.%
\footnote{%
  It can be argued that automatic generalization is acceptable,
  because ``for all'' is less problematic.
  One might still want a weaker version of the subformula property,
  saying that every type is either a subformula or related by generalization
  (and instantiation).
}
If we relax the recipe by including synthesis rules for $\unitexp$, integer literals
and similar constructs, we break the subformula property:
$\unitexp$ synthesizes $\unitty$ when the programmer never wrote $\unitty$.
However, a weaker---and still interesting---%
version of the property may hold,
since every type appearing in a derivation is either a subformula of a type in the conclusion
\emph{or} the ``obvious'' type of a literal constant.
That is, we can view $\unitexp$ as a request for the type $\unitty$.
Note that if we think of $\unitexp$ and integer literals as constants whose type is given
in a primordial context, so that instead of $\Gamma |- e : A$ we have
\[
   \unitexp : \unitty,\, 0 : \Int,\, -1 : \Int,\, 1 : \Int,\, -2 : \Int,\, \dots,\, \Gamma |- e : A
\]
then the full subformula property holds.
In effect, the author of the primordial context (the language designer)
has requested that $\unitexp$ and all the integer literals be permitted.
Other conveniences,
such as synthesizing the types of monomorphic functions,
can also be justified (with a little more work;
one must think of $\lambda$ as a sort of polymorphic constant).

In bidirectional systems, the goal in the checking judgment
can steer typing and avoid a measure of cleverness.
Thus, while the subformula property is not enjoyed by every imaginable bidirectional type system,
bidirectionality seems to make the property easier to achieve.

\subsection{Verifications and Uses}
\label{sec:verifications}

In the linear simply typed lambda calculus of \citet{Cervesato03},
typing is presented using two judgments:
a \emph{pre-canonical} judgment that
``validates precisely the well-typed terms\dots in $\eta$-long form''
and
a \emph{pre-atomic} judgment that
``handles intermediate stages of their [the terms'] construction''.
As suggested by the word ``validates'',
the pre-canonical judgment corresponds to checking,
and the pre-atomic judgment corresponds to synthesis.

Their notation differs from most of the early papers on bidirectional typing:
they write pre-canonical judgments as $M \Uparrow a$
and pre-atomic judgments as $M \downarrow a$,
which is almost exactly the reverse of (for example)
DML \citep{Xi99popl}, which used $\uparrow$ for synthesis
and $\downarrow$ for checking.
(Both notations are reasonable:
computer scientists usually write trees with the root at the top,
so Xi's arrows match the flow of type information through a syntax tree;
Gentzen put the root of a derivation tree at the bottom,
so Cervesato and Pfenning's arrows match the flow of type information through the typing derivation.)

This division into validation (checking) and handling intermediate values (synthesis) persists,
with different terminology,
in Frank Pfenning's teaching on verifications and uses:
A \emph{verification} of a proposition checks that it is true;
a \emph{use} of an assumption decomposes the assumed proposition.
We are not aware of a published paper describing verifications and uses,
but the idea appears in many of Pfenning's lecture notes.
The earliest seems to be \citet[p.\ 29]{Pfenning04:sequent-calculus},
with similar notation to \citet{Cervesato03}:
\begin{quote}
  $A \Uparrow$~~~~Proposition $A$ has a normal deduction, and
  \\
  $A \downarrow$~~~~~Proposition $A$ is extracted from a hypothesis.
\end{quote}
Later lecture notes introduce the terminology of \emph{verifications} and \emph{uses},
writing $\uparrow$ and $\downarrow$ respectively \citep{Pfenning09:harmony,Pfenning17:verifications}.
Verification is related \citep[p.\ 2]{Pfenning17:verifications} to 
``intercalation'' in proof search \citep{Sieg98}.

\section{Focusing, Polarized Type Theory, and Bidirectional Type Systems}
\label{sec:focusing}

A widespread folklore belief among researchers is that bidirectional
typing arises from \emph{polarized} formulations of logic. This belief
is natural, helpful, and (surprisingly) wrong.

\subsection{Bidirectional Typing and the Initial Cartesian Closed Category}

The naturalness of the connection can be seen from \Figureref{fig:stlc-norm},
which gives a bidirectional type system
that precisely characterizes $\beta$-normal, $\eta$-long terms.
The only necessary changes from \Figureref{fig:stlc} were:

\begin{itemize}
\item The annotation rule was removed. Since annotations are only
  required at $\beta$-redexes, the omission of this rule forces all
  typable terms to be $\beta$-normal.
\item The mode-switch rule from synthesis to checking is restricted
  to allow mode switches only at base type.
  This makes it impossible to partially apply a function in the checking mode:
  if $f : b \to b \to b$ and $x : b$, then $f\;x\;x$ checks
  but $f\;x$ does not.
  If a partial application is desired,
  it must be $\eta$-expanded to $\fun{y} f\;x\;y$. 
\end{itemize}

\begin{figure}[t]
  \centering
  
  \begin{grammar}
    Expressions
    & $e$
    & \bnfas
    &
        $
        x
        \bnfalt
        \Lam{x} e
        \bnfalt
        e\,e
        $
    \\
    Types
    & $A, B, C$
    & \bnfas
    &
       $
       b
       \bnfalt
       A -> A
       $
    \\
    Typing contexts
    & $\Gamma$
    & \bnfas
    &
       $
       \emptyctx
       \bnfalt
       \Gamma, x : A
       $
  \end{grammar}

      \judgbox{
        \Gamma |- e <= A
        \\
        \Gamma |- e => A
      }%
      {  ~\\[0.3ex]
        Under $\Gamma$, expression $e$ checks against type $A$
           \\[0.4ex]
           Under $\Gamma$, expression $e$ synthesizes type $A$
      }
      \vspace*{1.0ex}
      \begin{mathpar}
        \Infer{\SynVar}
            {(x : A) \in \Gamma}
            {\Gamma |- x => A}
        \\
        \Infer{\ChkSub}
            {
              \Gamma |- e => b
              \\
              b = B
            }
            {\Gamma |- e <= B}
        \\
        \Infer{\ChkArrIntro}
            {
              \Gamma, x : A_1 |- e <= A_2
            }
            {
              \Gamma |- (\Lam{x} e) <= A_1 -> A_2
            }
        \and
        \Infer{\SynArrElim}
            {
              \Gamma |- e_1 => A -> B
              \\
              \Gamma |- e_2 <= A
            }
            {\Gamma |- e_1\,e_2 => B}
      \end{mathpar}

  \caption{A bidirectional type system characterizing $\beta$-normal, $\eta$-long normal forms}
  \label{fig:stlc-norm}
\end{figure}

Together, these two restrictions ensure that only $\beta$-normal, $\eta$-long
terms typecheck.
Moreover, this characterization is easy to extend to products:
\begin{displaymath}
\begin{matrix}
\Rule{ }
     { \CHECK{\Gamma}{\unitexp}{1} }
&
\mbox{(No unit elimination rule)}
\\[1em]
\Rule{ \CHECK{\Gamma}{e_1}{A_1} \\
	   \CHECK{\Gamma}{e_2}{A_2} }
     { \CHECK{\Gamma}{(e_1, e_2)}{A_1 \times A_2} }
& 
\Rule{
  \synth{\Gamma}{e}{A_1 \times A_2}
  \\
  i \in \{1,2\}
}
{ \synth{\Gamma}{\pi_i(e)}{A_i} }
\end{matrix}
\end{displaymath}
This type system now characterizes normal forms in the STLC with
units and products.
Recall that the lambda calculus with units, pairs, and functions
is a syntax for the initial cartesian closed category,
when terms are quotiented by the $\beta\eta$ theory for each type~\citep{Lambek85}.

Since this bidirectional system requires $\beta$-normal, $\eta$-long terms,
we can see it as a calculus that presents the initial model of Cartesian closed categories
\emph{without} any quotienting.
Morphisms are well-typed terms,
and two morphisms are equal when they are $\alpha$-equivalent.

All that remains is to show that identities and composition are definable.
In the ordinary presentation of the initial CCC, a morphism
is a term with a free variable, and composition is substitution.
In the bidirectional system, however,
a morphism $A \to B$ is a checking term $\CHECK{x:A}{e}{B}$,
and substituting a checking term for a variable
does not preserve the $\beta$-normal, $\eta$-long property.
However, if we use \emph{hereditary substitution}
\citep{Pfenning01,Watkins03,Nanevski08}---%
a definition of substitution that also inspects the structure
of the term being substituted
and ``re-normalizes'' as it goes---%
then we restore the property that substitution preserves $\beta$-normal, $\eta$-long terms.

This means that the bidirectional type system constitutes a term model
for the initial CCC,
as follows:

\begin{enumerate}
\item The objects of the term are the types of the programming
  language.
\item Morphisms $X \to Y$ are terms $\CHECK{x : X}{e}{Y}$.
\item The identity morphism $\mathit{id} : X \to X$ is the
  $\eta$-expansion of the single free variable.
\item Composition of morphisms is given by hereditary substitution. 
\end{enumerate}

The usual presentation of the term model requires quotienting terms by $\beta\eta$-equivalence,
but the term model built from the bidirectional system has the property that
equality of morphisms is just $\alpha$-equivalence.
By interpreting a non-normal term into this category,
any two $\beta\eta$-equal terms will have the same denotation.

\subsection{Adding Problems with Sums}

This construction is so beautiful that it is essentially irresistible
to add sums to the language. Unfortunately, doing so introduces
numerous difficulties. These are most simply illustrated by using the
basic bidirectional recipe of \citet{Dunfield04:Tridirectional},
which yields an introduction and elimination rule for sum types as follows:
\begin{mathpar}
\Rule{ \CHECK{\Gamma}{e}{A_i} \\ i \in \{1,2\} }
	 { \CHECK{\Gamma}{\inj{i}{e}}{A_1 + A_2} }
\and
\Rule{ \synth{\Gamma}{e}{A_1 + A_2} \\ \CHECK{\Gamma, x_1:A_1}{e_1}{C} \\ \CHECK{\Gamma, x_2:A_2}{e_2}{C} }
	 { \CHECK{\Gamma}{\case{e}{x_1}{e_1}{x_2}{e_2}}{C} }
\end{mathpar}
These rules say that both the injection and case rules have
a checking conclusion, but that the scrutinee $e$ in the \textkw{case} must synthesize a sum type.
As we noted in \Sectionref{sec:recipe-intro-elim},
this imposes some restrictions on which terms are typeable.
For example,
because the rule for \textkw{case} has a checking conclusion,
we cannot use a \textkw{case} in function position without a type annotation:
\[
a:((b \to A) + (b \to A)), x:b \not\entails \case{a}{f}{f}{g}{g}\;x \From A
\]
Instead of applying an argument to a \keyword{case} expression of function type,
we must push the arguments into the branches:
\[
a:((b \to A) + (b \to A)), x:b \vdash \case{a}{f}{f\;x}{g}{g\;x} \From A
\]
If we intend to type only normal forms, this seems desirable:
these rules are prohibiting certain term forms that
correspond to commuting conversions of allowed terms.
The need for commuting conversions has never been popular with logicians:
witness Girard's lament~\citep[p. 79]{girard:proofs-and-types} that
``one tends to think that natural deduction should be modified to correct such atrocities.''
However, the simple bidirectional system does not completely
eliminate the need for commuting conversions.
For example, consider the term
\[
  \synth{f : b \to b, x:b+b}{f\;\big(\case{x}{y}{y}{z}{z}\big) }{b}
\]
This term is equivalent to the previous one by a commuting conversion,
but both terms are still typable.

Note that allowing the case form to synthesize a type (as
in \Sectionref{sec:recipe-intro-elim}) allows \emph{more} terms to
typecheck, which is the opposite of what we want (in this section, anyway).
In practice,
it can be difficult to support an unannotated case form which synthesizes its type.
Concretely, if the arm $e_1$ synthesizes the type $C_1$
and the arm $e_2$ synthesizes the type $C_2$,
we have to check that they are equal.
However, in the general case (e.g., in dependent type theory) equality is
\emph{relative to the context}, and the context is different in each
branch (with $\Gamma, x_1:A_1$ in one branch and $\Gamma, x_2:A_2$ in the other).
This is why dependent type theories like Coq end up
requiring \keyword{case} expressions to be annotated with a return type:
this resolves the problem by having the programmer solve it herself.

\subsection{A Polarized Type Theory}

At this point, we can make the following pair of observations:

\begin{enumerate}
\item
  The simple bidirectional system for
  the simply typed lambda calculus with products
  has the property that
  two terms are $\beta\eta$-equal if and only if they are the same:
  it fully characterizes $\beta\eta$-equality.
  
\item Adding sum types to the bidirectional system breaks this property:
  two terms equivalent up to (some) commuting conversions may both be typable. 
\end{enumerate}

To restore this property, two approaches come to mind.
The first approach is to find even more restrictive notions of
normal form which prohibit the commuting conversions.
We will not pursue this direction in this article,
but see \citet{scherer17} and \citet{ilik17} for examples of this approach.

The second approach is to find type theories in which the
commuting conversions \emph{no longer preserve equality}.
By adding (abstract) effects to the language, terms that used to be equivalent
can now be distinguished, ensuring that term equality once again
coincides with semantic equality. This is the key idea embodied in
what is variously called \emph{polarized type theory},
\emph{focalization}, or \emph{call-by-push-value}~\cite{LevyThesis}.

Now, we give a polarized type theory resembling those of \citet{Simmons14}
and \citet{EspiritoSanto17}.
Our main change is to adjust the proof term
assignment to look more familiar to functional programmers.
\begin{grammar}
{Positive types} & $P, Q$ & \bnfas &
$\unitty \bnfalt P \times Q \bnfalt P + Q \bnfalt \downshift{N}$
\\
{Negative types} & $N, M$ & \bnfas &
$P -> N \bnfalt \upshift{P}$
\\ 

{Values} & $v$ & \bnfas &
$u \bnfalt \unitexp \bnfalt (v,v) \bnfalt \inj{i}{v} \bnfalt \thunk{t}$
\\
{Spines} & $s$ & \bnfas &
$\cdot \bnfalt v\;s$
\\[1ex]
{Terms} & $t$ & \bnfas &
$\return{v}
\bnfalt \lambda{\overrightarrow{\arm{p_i}{t_i}}}
\bnfalt \Match{x \cdot s}{\overrightarrow{\arm{p_i}{t_i}}}$
\\
{Patterns} & $p$ & \bnfas &
$\unitexp \bnfalt (p,p') \bnfalt \inj{i}{p} \bnfalt \thunk{x}$
\\[1ex]
{Contexts} & $\Gamma,\Delta$ & \bnfas &
$\cdot \bnfalt \Gamma, x : N \bnfalt \Gamma, u : P$
\\
{Typing values} & & &
$\CHECK{\Gamma}{v}{P}$
\\
{Typing spines} & & &
$\spine{\Gamma}{s}{N}{M}$
\\
{Typing terms} & & &
$\CHECKn{\Gamma}{t}{N}$
\\
{Typing patterns} & & &
$\CHECKp{p:P}{\Delta}$
\end{grammar}

The key idea in polarized type theory is to divide types into two categories:
\emph{positive} types $P$ (sums, strict products, and suspended
computations)
and \emph{negative} types $N$
(basically, functions).
Positive types are eliminated by pattern matching,
and negative types are eliminated by supplying arguments.
Negative types can be embedded into positive types using the ``downshift'' type
$\downshift{N}$ (representing suspended computations);
positive types can be embedded into negative types
using the ``upshift'' $\upshift{P}$ (denoting computations producing $P$'s).

The semantics of call-by-push-value offer insight into the design of this calculus:
positive types correspond to objects of a category of values
(such as sets and functions),
and negative types correspond to objects of a category of computations
(objects are algebras of a signature for the computations,
and morphisms are algebra homomorphisms).
Upshift and downshift form an adjunction between values and computations,
and the monads familiar to functional programmers arise via the composite:
$T(P) \triangleq\; \downshift{\upshift{P}}$.

While this calculus arises from meditation upon invariants of proof theory,
its syntax is much closer to practical functional programming languages
than the pure typed lambda calculus,
including features like clausal definitions and pattern matching.
But the price we pay is a proliferation of judgments.
We usually end up introducing separate categories of
\emph{values} (for introducing positive types)
and \emph{spines} (argument lists for calling functions),
as well as \emph{terms}
(how to put values and spines together in computations,
as well as introducing negative types)
and \emph{patterns} (how to eliminate positive types).

Contexts have two kinds of variables,
$x:N$ for negative variables and $u:P$ for positive variables.

\subsubsection{Typing Values}
First, we give the typing rules for values.
As in the simple bidirectional recipe, we have a judgment $\CHECK{\Gamma}{v}{P}$
 for checking the type of positive values. 
\begin{mathpar}
    \Rule{}
    { \CHECK{\Gamma}{\unitexp}{\unitty} }
    \and 
    \Rule{ \CHECK{\Gamma}{v}{P} \\
      \CHECK{\Gamma}{v'}{Q} }
    { \CHECK{\Gamma}{(v,v')}{P \times Q} }
    \and
    \Rule{ \CHECK{\Gamma}{v}{P_i} \\ i \in \{1,2\} }
    { \CHECK{\Gamma}{\inj{i}{v}}{P_1 + P_2} }
    \\
\Rule{ \CHECKn{\Gamma}{t}{N} }
     { \CHECK{\Gamma}{\thunk{t}}{\downshift{N}} }
    \and
    \Rule{ (u : Q) \in \Gamma \\ P \equiv Q }
         { \CHECK{\Gamma}{u}{P} }
\end{mathpar}
The rules for units, pairs and sums are unchanged from the simple
bidirectional recipe. The rule for downshift says that if a term $t$
checks at a negative type $N$, then the thunked term $\thunk{t}$ will
check against the downshifted type $\downshift{N}$. Finally, a
variable $u$ checks at a type $P$ if the context says that $u$ has a type $Q$ equal to $P$.
(With subtyping, we would instead check that $Q$ is a subtype of $P$.) 

\subsubsection{Typing Spines}

Before we give the typing rules for all terms, we will give the rules deriving
the spine judgment $\spine{\Gamma}{s}{N}{M}$,
read ``if spine $s$ is applied to a head of type $N$, it will produce a result of type $M$''.
The type $N$ is an algorithmic input, and the type $M$ is an output. 
\begin{mathpar}
\Rule{  }
     { \spine{\Gamma}{\cdot}{N}{N} }
\and
\Rule{ \CHECK{\Gamma}{v}{P} \\ \spine{\Gamma}{s}{N}{M} }
     { \spine{\Gamma}{v\;s}{P \to N}{M} }
\end{mathpar}
The first rule says that an empty argument list does nothing to the
type of the head: the result is the same as the input. The second rule
says that a non-empty argument list $v\;s$ sends the function type
$P \to N$ to $M$, if $v$ is a value of type $P$ (i.e., a valid
argument to the function),
and $s$ is an argument list sending $N$ to $M$. 

\subsubsection{Typing Terms}

With values and spines in hand, we can talk about terms, in the term
typing judgment $\CHECKn{\Gamma}{t}{N}$, which checks that a term $t$
has the type $N$.
\begin{mathpar}
\Rule{ \CHECK{\Gamma}{v}{P} }
     { \CHECKn{\Gamma}{\return{v}}{\upshift{P}} }
~~~
\Rule
  {
    \arrayenvbl{
      \text{for all~}i < n.
      \\
      ~~~~\CHECKp{p_i:P}{\Delta_i}
      \\
      ~~~~\CHECKn{\Gamma, \Delta_i}{t_i}{N}
    }
  }
  {
    \CHECKn{\Gamma}{\lambda{\overrightarrow{\arm{p_i}{t_i}}^{i < n}}}{P \to N}
  }
~~~
\Rule{
  \arrayenvbl{
    (x:M) \in \Gamma
    \\
    \spine{\Gamma}{s}{M}{\upshift{Q}}
  }
    \\
    \arrayenvbl{
      \text{for all~}i < n.
      \\
      ~~~~\CHECKp{p_i:Q}{\Delta_i}
      \\
      ~~~~\CHECKn{\Gamma, \Delta_i}{t_i}{\upshift{P}}
    }  
  }
  {
    \CHECKn{\Gamma}{\Match{x \cdot s}{\overrightarrow{\arm{p_i}{t_i}}^{i < n}}}{\upshift{P}}
  }
\end{mathpar}
The rule for $\return{v}$ says that we embed a value $v$ of type
$P$ into the upshift type $\upshift{P}$ by immediately returning it.
Lambda abstractions are pattern-style---instead of a single binder $\lam{x}{t}$,
we give a list of patterns and branches
$\lambda{\overrightarrow{\arm{p_i}{t_i}}}$
to check at type $P -> N$.
As a result, we need a judgment $\CHECKp{p_i:P}{\Delta_i}$
giving the types of the bindings $\Delta_i$ of the pattern $p_i$,
and then we check each $t_i$ against the result type $N$.
Then we check each branch $t_i$ against the type $N$
in a context extended by $\Delta_i$. 

We face similar issues in the match expression
$\Match{x\cdot s}{\overrightarrow{\arm{p_i}{t_i}}}$.
First, it finds the variable $x$ in the context,
applies some arguments to it to find a value result of type $\upshift{Q}$,
and then pattern matches against type $Q$.
(In typical bidirectional systems, a synthesis judgment would type $x \cdot s$;
in this polarized system of normal forms, the synthesis judgment is absorbed into
the rule for match expressions.)
We check that the spine $s$ sends $M$ to the type $\upshift{Q}$,
and then check that the patterns $p_i$ yield variables $\Delta_i$ at the type $Q$,
we can check each $t_i$ against the type $\upshift{P}$.
Restricting the type at which we can match forces us to $\eta$-expand terms of function type.

Both lambdas and application/pattern-matching use the
judgment $\CHECKp{p:P}{\Delta}$ to find the types of the bindings.
The rules for these are straightforward:
\begin{mathpar}
\Rule{ }
     { \CHECKp{\thunk{x} : \downshift{\!N}}{x : N} }
\and
\Rule{ }
     { \CHECKp{\unit : \unitty}{\cdot} }
\and
\Rule{ \CHECKp{p_1 : P_1}{\Delta_1} \\ \CHECKp{p_2 : P_2}{\Delta_2} }
     { \CHECKp{(p_1,p_2) : (P_1 \times P_2)}{\Delta_1, \Delta_2} }
\and
\Rule{ \CHECKp{p:P_i}{\Delta} \\ i \in \{1,2\} }
     { \CHECKp{\inj{i}{p} : (P_1 + P_2)}{\Delta} } 
\and
  \Rule{ }
       {\CHECKp{u:P}{u : P}}
\end{mathpar}
Units yield no variables at type $\unitty$,
pair patterns $(p_1, p_2)$ return the variables of each component,
injections $\inj{i}{p}$ return the variables of the sub-pattern $p$,
and thunk patterns $\thunk{x}$ at type $\downshift{N}$
return that variable $x$ at type $N$.
Here, we omit a judgment to check whether a set of patterns is complete;
see \citet{Krishnaswami09} for a polarization-based approach. 

Value bindings $u : P$ bind a value of type $P$ to a variable $u$,
which allows values to be pattern-matched without being fully decomposed.
Hence, our calculus is weakly focused in the sense of \citet{Pfenning09}.

\subsection{Discussion}

Bidirectional typing integrates very nicely into a perspective based on polarized logic.
Indeed,
the ``application judgment'' in \citet{Dunfield13}
can be seen as a special case of the spine judgment,
and even systems not designed from an explicitly bidirectional perspective,
such as \citet{Serrano18},
have found it beneficial to work with entire argument lists.
In our view, this is because the spine judgment is well-moded,
making it easy to manage the flow of information through the argument list.

This close fit is not limited to argument lists, but also
extends to other features where functional languages go beyond kernel calculi,
such as pattern matching.
\citet{Krishnaswami09} shows how ML-style pattern matching
arises as the proof terms of a focused calculus,
and indeed the type system in that paper is bidirectional.
This system only covered simple types, but the approach scales well.
Our bidirectional type system for generalized algebraic data types \citep{Dunfield19}
goes much further,
including both universal and existential quantification,
GADTs, and pattern matching.
Nevertheless, it is built upon essentially the same idea of
applying bidirectional typing to a focused type theory.

However, despite the fact that the standard recipe of bidirectional typing
fits beautifully with focused logics, we should not lose sight of the fact
that the essence of bidirectional typing is the management of information flow.
Consequently, these techniques apply more broadly than polarized calculi,
as fundamental as they may be.
In \Sectionref{sec:variations}, we saw a number of systems with a different mode structure,
such as the mixed-direction types of \citet{Odersky01:ColoredLocal}, the strict type inference
of \citet{Chlipala05},
and the backwards bidirectional system of \citet{Zeilberger15}.
All of these reject the basic bidirectional recipe, but are undeniably bidirectional.

Thus, we would advise designers of new bidirectional systems to seek inspiration
from polarized type theory, but not to restrict themselves to it. 

\section{Other Applications of Bidirectional Typing}
\label{sec:applications}

\subsection{Dependent Types, Refinement Types, and Intersection Types}

The DML system \citep{Xi99popl,XiThesis} used bidirectional typing
because type inference for index refinements (a form of refinement type)
is undecidable.
DML followed a ``relaxed'' version of the Pfenning recipe
that allowed some rules that are not strictly necessary,
such as an introduction rule that synthesizes a type for
$\Pair{e_1}{e_2}$ if $e_1$ and $e_2$ synthesize.

The first datasort refinement system \citep{Freeman91}
used a form of type inference similar to abstract interpretation;
the later systems SML-CIDRE \citep{Davies00icfpIntersectionEffects,DaviesThesis}
and Stardust \citep{DunfieldThesis}
used bidirectional typing.
In SML-CIDRE, type inference was eschewed in favour of bidirectional typing:
type inference finds \emph{all} behaviours,
not only the \emph{intended} behaviours.
The type annotations in bidirectional typing,
especially when following the Pfenning recipe as SML-CIDRE did,
force programmers to write down the behaviours they intend.
In Stardust,
the decision to use bidrectional typing
was also motivated by the undecidability of type inference for index refinements.

In contextual modal type theory \citep{Nanevski08},
typing is bidirectional to make type checking decidable
in the presence of dependent types.
That theory is the main foundation for Beluga,
which is bidirectional for the same reason.
The original core of Beluga \citep{Pientka08:POPL,Pientka08:PPDP}
follows the Pfenning recipe,
but the full system \citep{Pientka13} %
extends the recipe, supporting both checking and synthesis for spines (lists of function arguments).

Bidirectional type checking was folklore among programming language
developers since the 1980s,
and was known and used by developers of proof assistants since the 1990s:
\citet{Coquand96:typechecking-dependent-types}
presents a type-checking algorithm for a small dependent type system
(dependent products, plus let-expressions),
in which ``type-checking'' ($\dots M \Rightarrow v$) 
and ``type inference'' ($\dots M \mapsto v$)
``inductively and simultaneously''.
Coquand's version of a subsumption rule \citep[p.\ 173, third part of definition]{Coquand96:typechecking-dependent-types}
says that $M$ checks against $v$ if $M$ synthesizes $w$ and $w$ is convertible to $v$.
Moreover, by removing the parts related to dependent typing,
we see that Coquand's rule for $\lambda$-application (the fifth part of his definition)
is essentially our standard rule:
to synthesize a type for $M_1\,M_2$, synthesize a type for $M_1$ and check $M_2$ against the domain of that type.
Strikingly, the Gofer code in the paper has functions \textvtt{checkExp} and \textvtt{inferExp}
that have the expected type signatures;
for example, \textvtt{checkExp} returns a boolean.

\citet{scherer2012irrelevance} give a bidirectional algorithm for
judgmental equality in dependent type theory.
They give a non-bidirectional specification of a dependent type system.
Then they give a bidirectional algorithmic system, which
defines where to do equality tests and leads to a nice algorithm for
how to do judgmental equality.
They prove soundness, completeness, and decidability of their algorithm
using two intricate logical relations; the combination of soundness and completeness
leads to decidability.
Judgmental equality is bidirectional: comparison of neutral terms synthesizes,
and comparison of normal terms checks.
This cleverly exploits the fact that
off-diagonal cases (e.g., atomic terms compared against a normal
terms) can be omitted, as an atomic term $t$ can only equal another
term $t'$ if $t'$ reduces to another term with the same head variable
as $t$.

\citet{McBride16} advocates a bidirectional system as the specification of a dependent type system.  Other bidirectional dependent type systems include
PiSigma \citep{Altenkirch10}, intended as a small core system for dependently typed
languages, and Zombie \citep{Sjoberg15}.
As with subtyping, conversion checking is not syntax-directed and is guided by
bidirectionality.
Taking the bidirectional system as the specification simplifies the
metatheory in some ways.
However, proving that an algorithmic conversion relation is an equivalence relation,
congruent for all the syntactic forms, still seems to require a sophisticated argument.

Intersection types, originally formulated in undecidable type assignment systems,
have motivated the use of bidirectional typing in several systems,
including refinement intersection types \citep{Dunfield04:Tridirectional}
and unrestricted intersection types \citep{Dunfield14,Oliveira16}
with polymorphism \citep{Alpuim17}.
Some of these systems also include union types.

\subsection{Gradual Typing}
\label{sec:gradual-typing}

Gradual typestate \citep{Wolff11}  %
uses bidirectional typing to structure the flow of information about access permissions,
specified in annotations.
Their language, descended from Featherweight Java, is imperative in flavour;
its expression forms are not easy to classify as introductions or eliminations,
making it hard to apply the Pfenning recipe.
Our discussion of reasoning by cases (step 3 in \Sectionref{sec:recipe-intro-elim})
carries over to their typing rules for \textkw{let}, which allow either (1) the body of the \textkw{let}
to synthesize, and hence the entire \textkw{let}, or (2) the body to be checked, based on a type
against which the entire \textkw{let} is to be checked.
(Our judgment form $\Gamma |- \cdots -| \Delta$ from \Sectionref{sec:polymorphism}
looks similar to the gradual typestate judgment $\Delta |- \cdots -| \Delta'$;
moreover, in both settings, the left-hand context is called the input context
and the right-hand context is called the output context.
However, the meaning is completely different.
In gradual typestate,
the output context describes the \emph{state} after running the subject expression,
so the output context often has different information than the input context.)

Gradual sum types \citep{Jafery17}
are formulated in a functional style,
so the Pfenning recipe works.
The subformula property ensures that uncertain types---%
connectives that relax the guarantees of static typing---%
appear only when the programmer asks for them.
In their subsumption rule (ChkCSub), the subtyping judgment is replaced by
\emph{directed consistency},
a relation that contains subtyping but also allows shifts between
more precise (less uncertain, more static) and less precise (more uncertain, less static)
types.

\citet{Xie18:ConsistentSubtypingForAll}
develop a gradual type system with
consistent subtyping (related to directed consistency)
and
higher-rank polymorphism.
Their bidirectional system closely follows \citet{Dunfield13},
discussed in \Sectionref{sec:complete-and-easy};
this approach leads to a subformula property that,
as in \citep{Jafery17},
ensures that the unknown type appears only by programmer request.

\subsection{Other Work}

\citet{Cicek19} define a relational type system
where each judgment has two subject terms (expressions):
$\Gamma |- e_1 \backsim e_2 : \tau$
relates the terms $e_1$ and $e_2$ at type $\tau$.
Their bidirectionalization follows
the Pfenning recipe in its original form,
for example,
their rule \textbf{alg-r-if}
for \textvtt{if} expressions has a checking conclusion.

\section{Historical Notes}
\label{sec:history}

Pierce and Turner's paper ``Local Type Inference''---%
which appeared as a technical report (\citeyear{Pierce97:LocalTypeInf}),
at POPL (\citeyear{Pierce98popl})
and in TOPLAS (\citeyear{Pierce00})---%
is the earliest frequently cited paper on bidirectional typing,
but Pierce noted that ``John Reynolds first acquainted us [BCP]
with the idea of bidirectional typechecking around 1988''.

That year also saw the first version of the report on Forsythe
\citep{Reynolds88:Forsythe},
where Reynolds noted that Forsythe's intersection types
would require some type information in the source program.
The second version of the report \citep[Appendix C]{Reynolds96:Forsythe}
describes an algorithm that
combines ``bottom-up typechecking'' and ``top-down checking'',
but the precise connection to bidirectional typing is not clear to us.

\citet{Lee98} present Algorithm $\mathcal{M}$, a type inference algorithm
used in some versions of Caml Light (the predecessor of OCaml).
In contrast to Algorithm $W$ \citep{Milner78}, if we reformulate Algorithm $\mathcal{M}$
as a set of typing rules, they are all checking rules: there is an input type $\rho$
to check against.
In some situations---%
for example, when typing a let-bound expression---%
the input type is a fresh unification variable,
but the input type often carries information.
They show that $\mathcal{M}$ can find type errors earlier than $W$,
which is consistent with the idea that bidirectional checking provides better error messages.

\citet{Dunfield04:Tridirectional} has two authors, but
the recipe was invented by Frank Pfenning,
so we call it the Pfenning recipe.

\section{Summary of Bidirectional Typing Notation}
\label{sec:notation}

\begin{table}[tb ]
  \centering

\newcommand{\SepLine}{\\ ~\\[-1ex]}
{\runonfontsz{9pt}
\begin{tabular}[t]{rccllllllllllll}
  & checks against & synthesizes
\\[6pt]
\citet{Coquand96:typechecking-dependent-types} &
$\Rightarrow$ &
$\mapsto$
\SepLine
\citet{Pierce00} &
$\Piercechk$ &
$\Piercesyn$
\SepLine
\citet{Xi99popl};
\citet{Davies00icfpIntersectionEffects};
&
$\downarrow$ &
$\uparrow$ %
\\
\citet{Dunfield04:Tridirectional}; &
\\
\citet{Polikarpova16}; \citet{Cicek19} &
\SepLine
\citet{Chlipala05};
\citet{Pottier06:stratified};
&
$\Downarrow$ &
$\Uparrow$ %
\\[2pt]
\citet{Dunfield09}
\\[2pt]
\citet{PeytonJones07}
&
$\entails_{\Downarrow}$ &
$\entails_{\Uparrow}$ %
\SepLine
\citet{DaviesThesis} &
$\DaviesThesischk$ &
$\DaviesThesissyn$
\SepLine
\tabularenvr{%
\citet{Nanevski08};
\\ \citet{Pientka08:POPL},
\citet{Pientka08:PPDP};
\\ \citet{Wolff11};
\citet{Dunfield12,Dunfield14};
\\
\citet{Dunfield13};
\citet{Dunfield15};  %
\\
\citet{Oliveira16};
\citet{Jafery17};
\\
\citet{Xie18:LetArgumentsGoFirst,Xie18:ConsistentSubtypingForAll}
} &
$\Leftarrow$ &
$\Rightarrow$ %
\SepLine
\citet{McBride16} &
$A \ni e$  &
$e \in A$
\SepLine                 
\citet{Lindley17}  & $e : A$ & $e \Rightarrow A$
\end{tabular}
}

\bigskip

  \caption{Historical and recent notation}
  \label{tab:notation}
\end{table}

\Tableref{tab:notation} summarizes some of the symbols that have been used
to denote checking and synthesis.
Until about 2008, most authors used vertical arrows
($\downarrow$ for checking and $\uparrow$ for synthesis),
though \citet{Pierce00} used $\Piercechk$ for checking and $\Piercesyn$ for synthesis.
The arrows were meant to represent information flow,
but vertical arrows are unclear because
syntax trees and derivation trees put the root at opposite ends:
does $e \uparrow A$ mean that the type flows from a leaf of a syntax tree
(at the bottom, away from the root),
or from the conclusion of a derivation tree?

Horizontal arrows avoid this confusion:
nearly all authors write the subject term to the left of the type in a judgment,
so $e => A$ means that the type is flowing from the term
and $e <= A$ means that the type is flowing ``into'' the term.

The $\ni$/$\in$ notation, used by \citet{McBride16},
has the advantage that information always flows left to right.

\section{Conclusion}

Bidirectional typing is usually straightforward to implement.
However, while the bidirectional approach allows us to prove
soundness, completeness and decidability of state-of-the-art typing algorithms,
the proofs are often extremely involved.
Moreover, the proofs are about type systems that focus on a few features of
research interest.
Since realistic programming languages combine many typing features,
doing the metatheory for a full-scale type system seems intractable using current techniques.
A key challenge for future research is to discover techniques to simplify the proofs of soundness, completeness, and decidability, enabling researchers to 
prove these key properties for complete programming languages.

Another future research direction is
that polarity, focusing and call-by-push-value seem to fit nicely with bidirectional typing, but are not the same as bidirectional typing.
We would like to understand the actual relationship between these ideas,
making it easier to integrate advances in proof theory into language design.

Designers of bidirectional type systems often employ information
flow that goes beyond checking (where the entire type is input)
and synthesis (where the entire type is output).
We would like to have a better understanding of how such
information flow interacts with core metatheoretic properties,
such as substitution principles, in order to arrive at broadly
applicable design principles for type system design.

\bibliography{survey}

\end{document}

